\documentclass[letter, 12pt]{article}\usepackage[]{graphicx}\usepackage[dvipsnames]{xcolor}
\makeatletter
\def\maxwidth{ %
  \ifdim\Gin@nat@width>\linewidth
    \linewidth
  \else
    \Gin@nat@width
  \fi
}
\makeatother

\definecolor{fgcolor}{rgb}{0.345, 0.345, 0.345}
\newcommand{\hlnum}[1]{\textcolor[rgb]{0.686,0.059,0.569}{#1}}%
\newcommand{\hlcom}[1]{\textcolor[rgb]{0.678,0.584,0.686}{\textit{#1}}}%
\newcommand{\hlopt}[1]{\textcolor[rgb]{0,0,0}{#1}}%
\newcommand{\hldef}[1]{\textcolor[rgb]{0.345,0.345,0.345}{#1}}%
\newcommand{\hlkwb}[1]{\textcolor[rgb]{0.69,0.353,0.396}{#1}}%
\newcommand{\hlkwc}[1]{\textcolor[rgb]{0.333,0.667,0.333}{#1}}%
\newcommand{\hlkwd}[1]{\textcolor[rgb]{0.737,0.353,0.396}{\textbf{#1}}}%

\usepackage{framed}
\makeatletter
\newenvironment{kframe}{%
 \def\at@end@of@kframe{}%
 \ifinner\ifhmode%
  \def\at@end@of@kframe{\end{minipage}}%
  \begin{minipage}{\columnwidth}%
 \fi\fi%
 \def\FrameCommand##1{\hskip\@totalleftmargin \hskip-\fboxsep
 \colorbox{shadecolor}{##1}\hskip-\fboxsep
     \hskip-\linewidth \hskip-\@totalleftmargin \hskip\columnwidth}%
 \MakeFramed {\advance\hsize-\width
   \@totalleftmargin\z@ \linewidth\hsize
   \@setminipage}}%
 {\par\unskip\endMakeFramed%
 \at@end@of@kframe}
\makeatother

\definecolor{shadecolor}{rgb}{.97, .97, .97}
\definecolor{messagecolor}{rgb}{0, 0, 0}
\definecolor{warningcolor}{rgb}{1, 0, 1}
\definecolor{errorcolor}{rgb}{1, 0, 0}
\newenvironment{knitrout}{}{} 

\usepackage{alltt}
\usepackage{amsmath, amssymb} 
\usepackage{array} 
\usepackage{doi} 
\usepackage[round]{natbib} 
\usepackage{multirow} 
\usepackage{booktabs} 
\usepackage[title]{appendix} 
\usepackage{nameref} 
\usepackage[dvipsnames]{xcolor}
\usepackage[doublespacing]{setspace} 
\usepackage{helvet}
\usepackage{mathpazo}
\usepackage{sectsty} 
\allsectionsfont{\sffamily} 
\usepackage[labelfont={bf,sf},font=small]{caption} 
\usepackage{orcidlink} 
\usepackage{pdflscape} 
\usepackage{afterpage} 
\usepackage{geometry}
\title{\vspace{-3em} \textbf{\textsf{
      Stabilizing Thompson Sampling with Null Hypothesis Bayesian Response-Adaptive Randomization
}}}
\author{
  Samuel Pawel \orcidlink{0000-0003-2779-320X}
  \and
  Leonhard Held \orcidlink{0000-0002-8686-5325}
}
\date{
  Epidemiology, Biostatistics and Prevention Institute (EBPI) \\
  Center for Reproducible Science and Research Synthesis (CRS) \\
  University of Zurich \\
  \texttt{\{samuel.pawel,leonhard.held\}@uzh.ch} \\[1ex]
  October 15, 2026
}

\usepackage{hyperref}
\hypersetup{
  bookmarksopen=true,
  breaklinks=true,
  colorlinks=true,
  linkcolor=black,
  anchorcolor=black,
  citecolor=blue,
  urlcolor=blue,
}

\IfFileExists{upquote.sty}{\usepackage{upquote}}{}
\begin{document}

\begin{onehalfspacing}
\maketitle
\begin{abstract}
  \noindent Response-adaptive randomization (RAR) methods can be used to adapt randomization probabilities based on accumulating data, aiming to increase the probability of allocating patients to effective treatments. A popular RAR method is Thompson sampling, which randomizes patients proportionally to the Bayesian posterior probability that each treatment is the most effective. However, its high variability can also increase the risk of assigning patients to inferior treatments and lead to inferential problems such as confidence interval undercoverage. We propose a principled method based on Bayesian hypothesis testing to address these issues: We introduce a null hypothesis postulating equal effectiveness of treatments. Bayesian model averaging then induces shrinkage toward equal randomization probabilities, with the degree of shrinkage controlled by the prior probability of the null hypothesis. Equal randomization and Thompson sampling arise as special cases when the prior probability is one or zero, respectively. A simulation study demonstrates that the method can mitigate issues with Thompson sampling and has comparable statistical properties to Thompson sampling with common ad hoc modifications such as power transformation and probability capping. Under the null hypothesis and a normal model, the randomization probabilities are shown to  converge asymptotically to equal randomization, unlike those of Thompson sampling. We implement the method in the free and open-source R package \texttt{brar}, enabling experimenters to easily perform null hypothesis Bayesian RAR and support more effective randomization of patients.
\end{abstract}
\end{onehalfspacing}

\section{Introduction}
Response-adaptive randomization (RAR) methods randomly allocate patients to
treatments based on accumulating data \citep{Hu2006, Thall2007,
  berry2010bayesian, Robertson2023}. A popular approach is Thompson sampling
\citep{Thompson1933}, which randomizes patients proportionally to the Bayesian
posterior probability that each treatment is the most effective. Such RAR
methods are attractive as they balance gathering information on treatment
effectiveness and assigning subjects to effective treatments.

Despite its benefits, Thompson sampling can also introduce inferential problems
such as inflated type I error rates or bias in effect estimation. Furthermore,
although Thompson sampling allocates patients to the more effective treatment on
average, it may also increase the probability of allocating patients to inferior
treatments compared to equal randomization, particularly when treatment effects
are small. This can undermine the method's intended objective and raise ethical
concerns \citep{Thall2015, Robertson2023}. Consequently, there has been interest
in modifying Thompson sampling to address these issues.

One popular modification uses the randomization probability $\pi = p^c / \{p^c +
(1 - p)^c\}$, where $p$ is the posterior probability that the experimental
treatment is more effective than the control, and $c$ is a tuning parameter.
Setting $c = 1$ produces Thompson sampling while $c < 1$ reduces variability.
Another approach is to cap randomization probabilities, for instance, setting
them to 10\% or 90\% if the method assigns more extreme probabilities. Finally,
``burn-in'' periods at the start of the study are commonly used, during which
patients are randomized with equal probabilities to mitigate high variability
\citep{Wathen2017}.

Several simulation studies have shown that appropriately modified Thompson
sampling can overcome the limitations of the ordinary formulation
\citep[e.g.,][]{Du2017, Viele2019, Viele2020, Kim2021, Tang2025}. However, such
ad hoc modifications conflict with principles of coherent Bayesian learning. For
example, a capped posterior no longer corresponds to an actual posterior and
cannot serve as a genuine prior for future data. This raises the question of
whether a principled Bayesian RAR method can mitigate Thompson sampling's
variability, and if so, how it relates to ad hoc modifications.

\begin{sloppypar}
Principled Bayesian alternatives to Thompson sampling have also been proposed
outside clinical trials, particularly in the multi-armed bandit literature. For
instance, RAR based on the Gittins index prioritizes treatments with the
greatest expected reward \citep{Gittins2011}, and the Bayesian upper confidence
bound method allocates treatments based on posterior quantiles and offers
theoretical performance guarantees \citep{Kaufmann2012}. These methods are
primarily used in online decision problems but rarely in clinical trials, where
additional constraints such as type I error rate control are central
\citep{Villar2015}, and where their often deterministic allocation rules can
increase the risk of selection bias and preclude randomization-based inference
methods \citep{Rosenberger2015, Bowden2017}.
\end{sloppypar}

In this paper, we propose a novel RAR method that reduces variability in a
coherent Bayesian manner, which we term ``\emph{null hypothesis Bayesian RAR}''.
The idea is to introduce a null hypothesis postulating that treatments are
equally effective. This is often equivalent to using a ``spike-and-slab'' prior
for the treatment effect parameter,
which is a mixture of a point mass at equal effectiveness and a probability
density elsewhere \citep[][Section~5.5.4]{Spiegelhalter2004b}. The
prior probability of the null hypothesis determines the mixture weight. Setting
it to one produces equal randomization, whereas setting it to zero produces
Thompson sampling. The method thus interpolates between equal randomization and
Thompson sampling in a coherent Bayesian way. As a by-product, posterior
probabilities and Bayes factors are also obtained. These can be used to monitor
evidence of treatment effectiveness in a manner that is aligned with the
randomization probabilities.

In the following Section \ref{sec:general-theory} we introduce the general idea
of the method, followed by tailoring it to normal (Section \ref{sec:normal}) and
binary outcomes (Section \ref{sec:binary}). Section \ref{sec:applications}
illustrates the method on data from the ECMO trial, followed by evaluating its
properties in a simulation study (Section \ref{sec:simulation}). Concluding
remarks are given in Section \ref{sec:discussion}. The supplement provides
further technical details, additional simulation results, and illustrates our R
package \texttt{brar}.

\section{Null hypothesis Bayesian RAR}
\label{sec:general-theory}
Assume that we have observed data $y$ and want to use these to randomize a
future patient. Suppose there is one control and one treatment group, and
consider the hypotheses:
\begin{align*}
  &H_{-} \colon \text{Treatment is less effective than control}& \\
  &H_{0} \colon \text{Treatment and control are equally effective}& \\
  &H_{+} \colon \text{Treatment is more effective than control.}&
\end{align*}
How exactly these are translated into statistical hypotheses about parameters
depends on the data model, but often relates to an effect size parameter being
less than, equal to, or greater than zero. If there is no control group, the method can
still be used with the control replaced by a reference group. This hypothesis
setup is similar to \citet{Cai2013} but with the addition of $H_0$, which
introduces a different behavior of the resulting randomization probabilities.

The Bayesian posterior probability of hypothesis $H_i \in \{H_{-}, H_0, H_{+}\}$
is then
\begin{align}
  \label{eq:posterior}
  \Pr(H_i \,\vert\, y)
  = \frac{p(y \,\vert\, H_i) \Pr(H_i)}{\sum\limits_{j \in \{-, 0, +\}} p(y \,\vert\, H_j) \Pr(H_j)}
  = 1 \, \bigg / \, \bigg\{\sum_{j \in \{-, 0, +\}} \mathrm{BF}_{ji}(y) \, \frac{\Pr(H_{j})}{\Pr(H_i)} \bigg\}
\end{align}
where $\Pr(H_i)$ is the prior probability of $H_i$ (with $\sum_{i \in \{-, 0,
  +\}} \Pr(H_{i}) = 1$), $p(y \,\vert\, H_i) = \int_{\Theta} p(y \mid \theta) \,
p(\theta \,\vert\, H_i) \, \mathrm{d} \theta$ is the marginal likelihood of the data
$y$ under $H_i$ obtained from integrating the likelihood $p(y \,\vert\, \theta)$ with
respect to the prior distribution $p(\theta \,\vert\, H_i)$ of the model parameters
$\theta \in \Theta$ under $H_i$, and
\begin{align}
  \label{eq:BF}
\mathrm{BF}_{ji}(y)
= \frac{\Pr(H_j \,\vert\, y)}{\Pr(H_i \,\vert\, y)} \, \bigg / \, \frac{\Pr(H_j)}{\Pr(H_i)}
= \frac{p(y \,\vert\, H_j)}{p(y \,\vert\, H_i)}
\end{align}
is the Bayes factor contrasting $H_j$ to $H_i$ \citep{Kass1995}. The Bayes
factor~\eqref{eq:BF} is the updating factor of the prior odds of $H_j$ versus $H_i$
to the corresponding posterior odds (first equality), which is equivalent to the
ratio of marginal likelihoods of the data under $H_j$ and $H_i$ (second
equality). The posterior probabilities~\eqref{eq:posterior} can thus be computed
from the marginal likelihoods of the data under each considered hypothesis along
with their prior probabilities, or from the Bayes factors and prior odds
relative to some reference hypothesis.

\begin{figure}[!htb]
\begin{knitrout}
\definecolor{shadecolor}{rgb}{0.969, 0.969, 0.969}\color{fgcolor}

{\centering \includegraphics[width=\maxwidth]{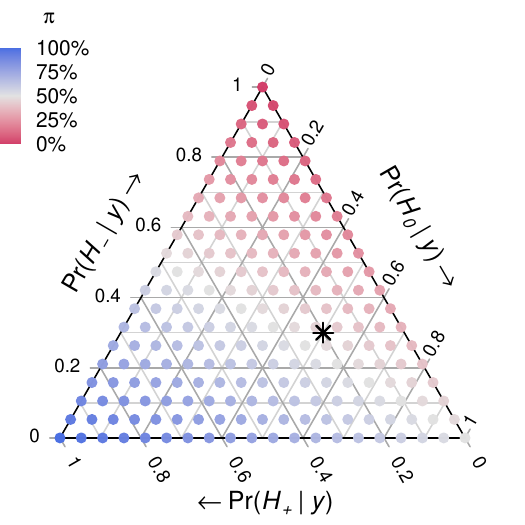} 

}

\end{knitrout}
\caption{Ternary plot where the color of each point depicts the treatment
  randomization probability $\pi$ for a particular combination of posterior
  probabilities of $H_{+}$ (bottom axis), $H_{-}$ (left axis), and $H_{0}$
  (right axis). For example, the black asterisk denotes the combination where
  $\Pr(H_{+} \,\vert\, y) = 0.2$, $\Pr(H_{-} \,\vert\, y) = 0.3$, and $\Pr(H_0 \,\vert\, y) =
  0.5$ with corresponding randomization probability $\pi = 45\%$.}
\label{fig:ternary}
\end{figure}

Regardless of which way they are computed, the question is how to translate
posterior probabilities into randomization probabilities. We propose to
randomize a future patient to the treatment group with probability
\begin{align}
  \label{eq:prand}
  \pi =  \Pr(H_{+} \,\vert\, y) + \Pr(H_0 \,\vert\, y) \, / \,  2,
\end{align}
while the probability to randomize to the control group is consequently $1 - \pi
= \Pr(H_{-} \,\vert\, y) + \Pr(H_0 \,\vert\, y) \, / \, 2$. Figure~\ref{fig:ternary} shows
$\pi$ for different combinations of posterior probabilities. We can see that
$\pi$ shrinks towards 50\% as the posterior probability of the null hypothesis
$\Pr(H_0 \,\vert\, y)$ increases. For example, if $\Pr(H_{+} \,\vert\, y) = 0.2$,
$\Pr(H_{-} \,\vert\, y) = 0.3$, and $\Pr(H_0 \,\vert\, y) = 0.5$, as indicated by the
black asterisk, the randomization probability is $\pi = 0.2 + 0.5/2 = 45\%$.

This scheme can be motivated as a Bayesian hypothesis-averaged randomization
probability where the hypothesis-specific treatment randomization probabilities
are $0\%$, $50\%$, and $100\%$, under $H_{-}$, $H_{0}$, and $H_{+}$,
respectively. That is, if we knew that $H_{+}$ or $H_{-}$ is true, we should
assign the next patient to the treatment or control group, respectively, to
maximize patient benefit. In contrast, if we knew that $H_{0}$ is true,
randomizing with $\pi = 50\%$ seems sensible.

The randomization scheme~\eqref{eq:prand} reduces to Thompson sampling when the
prior probability of $H_0$ is set to zero since then $\Pr(H_0 \,\vert\, y) = 0$
regardless of the data, and consequently $\pi = \Pr(H_{+} \,\vert\, y)$. However, the
scheme induces shrinkage toward $\pi = 50\%$ otherwise. In the most extreme case
when $\Pr(H_0) = 1$, equal randomization is obtained as then $\Pr(H_{0} \,\vert\, y)
= 1$ regardless of the data, and consequently $\pi = 50\%$. The scheme thus
interpolates between Thompson sampling and equal randomization in a coherent
Bayesian way.

\subsection{More than two groups}
\label{sec:multiple-treatments}
Suppose there are $K > 1$ treatment groups in addition to the control group. In
this case, we may modify the procedure and consider the hypotheses:
\begin{align*}
  &H_{-} \colon \text{All treatments are less effective than control}& \\
  &H_{0} \colon \text{All treatments are equally effective as control}& \\
  &H_{+i} \colon \text{Treatment $i$ is more effective than control and all other treatments}&
\end{align*}
for $i = 1, \dots, K,$ which span all effect ordering possibilities. Posterior
hypothesis probabilities can still be computed from~\eqref{eq:posterior} but
summing over all hypotheses (i.e., $j \in \{-, 0, +1, \dots, +K\}$). These can
then be translated into treatment randomization probabilities
\begin{align}
  \label{eq:multirand}
  \pi_i = \Pr(H_{+i} \,\vert\, y) + \Pr(H_0 \,\vert\, y) \, / \,  (K + 1)
\end{align}
and control randomization probability $1 - \sum_{i = 1}^K \pi_i = \Pr(H_{-} \,\vert\,
y) + \Pr(H_0 \,\vert\, y) \, / \, (K + 1)$.

Also in the multi-treatment case, the randomization probabilities shrink towards
equal randomization $\pi_i = 1/(K + 1)$ by introducing the null hypothesis
$H_0$. Thompson sampling and equal randomization are again obtained by setting
$\Pr(H_0) = 0$ and $\Pr(H_0) = 1$, respectively, as this leads to $\Pr(H_0 \,\vert\,
y) = 0$ and $\Pr(H_0 \,\vert\, y) = 1$ for any data $y$.

In this paper, we focus on scheme \eqref{eq:multirand}, which induces shrinkage
toward equal randomization but alternative schemes are also possible. It may be
desired to shrink to different ``baseline'' randomization probabilities. This
can be achieved by modifying the factor of $\Pr(H_0 \,\vert\, y)$
in~\eqref{eq:multirand} from $1/(K + 1)$ to the desired baseline randomization
probability. For example, the aim may be to test all $K$ pairwise differences
between the treatments and the common control using Dunnett's test. A square
root allocation rule (with $\sqrt{K}$ times as many patients in the control group as in
each treatment group) is commonly used to maximize the power of the test
assuming equal outcome variances across groups \citep[p.156]{Kieser2020}. We may
therefore use $\pi_i = \Pr(H_{+i} \,\vert\, y) + \Pr(H_0 \,\vert\, y) \, / \, (K
+ \sqrt{K})$. These shrink RAR probabilities towards Dunnett-type randomization
probabilities, $\pi_i = 1/(K + \sqrt{K})$ for treatment $i = 1, \dots, K$, and
$1 - \sum_{i = 1}^K \pi_i = \sqrt{K}/(K + \sqrt{K})$ for control. However,
Dunnett-type or other alternative baseline probabilities may seem unnatural from
a Bayesian hypothesis-averaging perspective, since knowing $H_0$ to be true,
there is no power to maximize and equal randomization is a more natural choice.

\subsection{Asymptotic randomization probabilities}
\label{sec:asymptotic}

It is of interest to understand how RAR probabilities behave as data accumulate.
Writing $H_1 = \bigcup_{j \neq 0} H_j$ for the alternative hypothesis, the
randomization probability to treatment group $i$ can be expressed as the mixture
\begin{align}
  \label{eq:mixture}
  \pi_i = w \times \underbrace{\Pr(H_{+i} \mid y, H_1)}_{\text{Thompson sampling}}
  + (1 - w) \times \kern-1em \underbrace{1/(K + 1)}_{\text{equal randomization}}
\end{align}
with weight $w = 1/[1 + \mathrm{BF}_{01}(y) \times \Pr(H_0) / \{1 - \Pr(H_0)\}]$
governed by the Bayes factor $\mathrm{BF}_{01}(y)$ contrasting $H_0$ to $H_1$.
When the data favor $H_1$, $\mathrm{BF}_{01}(y)$ is small, the weight $w$ is
close to one, and the randomization probabilities lie close to Thompson
sampling. When the data favor $H_0$, $\mathrm{BF}_{01}(y)$ is large, the weight
$w$ is close to zero, and the randomization probabilities lie close to equal
randomization.

The latter limiting behavior can be established formally under $H_0$. Section~A
of the supplement shows that the Bayes factor is almost surely bounded away from
zero by Ville's inequality because the inverse Bayes factor is a test martingale
for $H_0$ \citep{Shafer2011}. This in turn ensures that every group is sampled
infinitely often, so that estimators which are consistent under non-adaptive
sampling remain consistent under null hypothesis Bayesian RAR \citep{Melfi2000}.
The supplement further shows that under the normal model (introduced in the next
Section~\ref{sec:normal}) the Bayes factor diverges to infinity, so that the
randomization probabilities converge in probability to equal randomization if
$\Pr(H_0) > 0$. This is an important difference from Thompson sampling, whose
randomization probabilities do not converge under $H_0$ but remain random
\citep{Deliu2025}. The following Section~\ref{sec:normal} also demonstrates this
convergence empirically through simulation, and likewise provides empirical
support that the randomization probability of the superior treatment converges
to one under $H_1$.

\section{Null hypothesis Bayesian RAR under normality}
\label{sec:normal}
Suppose that the data are summarized by $y = \{\hat{\theta}, \sigma_n\}$, where
$\hat{\theta}$ is an estimate of the true treatment effect $\theta$, and
$\sigma_n$ is the (known) standard error of the estimate. This does not mean
that the raw data, from which the estimate is computed, need to be normally
distributed. For example, $\hat{\theta}$ could be an estimated (standardized)
mean difference, log odds/risk/rate/hazard ratio, risk difference, or regression
coefficient. The standard error is often of the form $\sigma_n =
\lambda/\sqrt{n}$, where $n$ is the effective sample size and $\lambda$ the
standard deviation of one effective observation. For example, for a mean
difference, the effective sample size is $n = 1/(1/n_C + 1/n_1)$ where $n_C$ and
$n_1$ are the sample sizes in the control and treatment group, respectively,
with $\lambda$ the standard deviation of one observation. For a log hazard
ratio, the effective sample size is the number of events with $\lambda = 2$.
Assume that the estimate is (at least approximately) normally distributed around
$\theta$ with its variance $\sigma^2_n$ assumed to be known, i.e., $\hat{\theta}
\,\vert\, \theta \sim \mathrm{N}(\theta, \sigma^2_n)$.

If the effect $\theta$ is oriented such that a positive effect indicates
treatment benefit, the hypotheses from Section~\ref{sec:general-theory}
translate into $H_{-} \colon \theta < 0$ vs. $H_{0} \colon \theta = 0$ vs.
$H_{+} \colon \theta > 0$. The null hypothesis $H_0$ is a simple hypothesis with
no free parameters, or equivalently, a point (Dirac) prior at zero. The $H_{-}$
and $H_{+}$ hypotheses are composite hypotheses and require prior distributions
for $\theta$. A natural choice is a $\theta \sim \mathrm{N}(\mu, \tau^2)$
distribution whose support is truncated to the negative and positive side,
respectively. In the absence of prior knowledge, it is sensible to center the
prior on zero ($\mu = 0$) to represent clinical equipoise \citep{Freedman1987},
giving equal probability to harmful and beneficial effects. Since $H_0$ is a
point hypothesis, the prior of $\theta$ under $H_{-}$ and $H_{+}$ must be proper
($\tau < \infty$), as otherwise the posterior probability of $H_0$ is 1 for any
data $y$. The value of the prior standard deviation $\tau$ must therefore be
chosen so that the prior is not too vague \citep{Kass1995}.

\begin{figure}[!htb]
\begin{knitrout}
\definecolor{shadecolor}{rgb}{0.969, 0.969, 0.969}\color{fgcolor}

{\centering \includegraphics[width=\maxwidth]{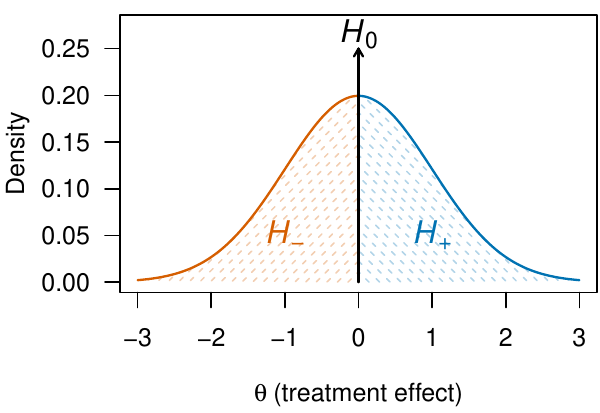} 

}

\end{knitrout}
\caption{Illustration of spike-and-slab prior for the effect $\theta$. A point
  prior at 0 is assumed under $H_0$. A normal prior $\theta \sim
  \mathrm{N}(0, 1)$ with support truncated to the positive
  or negative side is assumed under $H_{+}$ and $H_{-}$, respectively. These
  priors are averaged assuming prior hypothesis probabilities $\Pr(H_0) =
  0.5$,
  $\Pr(H_{+}) =
  0.25$, and
  $\Pr(H_{-}) = 
  0.25$.}
\label{fig:spike-slab}
\end{figure}

The prior probability of the null hypothesis $\Pr(H_0)$ represents the a priori
plausibility of an absent effect, and also acts as a tuning parameter that
controls the variability of RAR. An intuitive default is $\Pr(H_0) = 0.5$,
representing equipoise between absent and present effects \citep{Johnson2013b}.
For a given $\Pr(H_0)$, it is then natural to set the prior probabilities of
$H_{+}$ and $H_{-}$ to $\Pr(H_{+}) = \{1 - \Pr(H_0)\} \times \Phi(\mu/\tau)$ and
$\Pr(H_{-}) = \{1 - \Pr(H_0)\} \times \Phi(-\mu/\tau)$, with $\Phi(\cdot)$ the
standard normal CDF, so that the prior distribution averaged over $H_{+}$ and
$H_{-}$ is again the $\mathrm{N}(\mu, \tau^2)$ distribution. For example,
specifying $\Pr(H_0) = 0.5$ and a zero-centered prior, we obtain $\Pr(H_{+}) =
\Pr(H_{-}) = 0.5 \times 0.5 = 0.25.$ Averaging the prior over all hypotheses
then leads to a spike-and-slab prior, as illustrated in
Figure~\ref{fig:spike-slab}.

\begin{sloppypar}
With the priors specified, we can compute the marginal likelihood of the
observed effect estimate under each hypothesis and obtain posterior
probabilities~\eqref{eq:posterior}. In the normal likelihood and prior
framework, all are available in closed form. Denoting by $\mathrm{N}(x \,\vert\, m,
v)$ the normal density function with mean $m$ and variance $v$ evaluated at $x$,
the marginal likelihoods are
\begin{align*}
  p(\hat{\theta} \,\vert\, H_{-}, n)
  &= \mathrm{N}(\hat{\theta} \,\vert\, \mu, \sigma^2_n + \tau^2) \times
  \Phi(-\mu_*/\tau_*)  \, / \, \Phi(-\mu/\tau)
  \\
  p(\hat{\theta} \,\vert\, H_0, n)
  &= \mathrm{N}(\hat{\theta} \,\vert\, 0, \sigma^2_n)
  \\
  p(\hat{\theta} \,\vert\, H_{+}, n)
  &= \mathrm{N}(\hat{\theta} \,\vert\, \mu, \sigma^2_n + \tau^2) \times
  \Phi(\mu_*/\tau_*) \, / \, \Phi(\mu/\tau)
\end{align*}
with posterior mean and variance $\mu_* = (\hat{\theta}/\sigma^2_n + \mu/\tau^2)
\tau^2_*$ and $\tau_*^2 = 1/(1/\sigma^2_n + 1/\tau^2)$, where the standard error
$\sigma_n$ depends on the current sample size $n$.
Taking ratios of marginal likelihoods leads to the Bayes factors
\begin{align*}
  \mathrm{BF}_{+0}(\hat{\theta})
  &= \exp\left[-\frac{1}{2}\left\{\frac{(\hat{\theta} - \mu)^2}{\sigma^2_n + \tau^2} -
    \frac{\hat{\theta}^2}{\sigma^2_n} \right\}\right] \, \times \,
  \frac{\Phi(\mu_*/\tau_*)}{\Phi(\mu/\tau)} \, \bigg / \, \sqrt{1 + \frac{\tau^2}{\sigma^2_n}} \\
  \mathrm{BF}_{+-}(\hat{\theta})
  &= \frac{\Phi(\mu_*/\tau_*)}{\Phi(\mu/\tau)} \, \bigg / \, \frac{\Phi(-\mu_*/\tau_*)}{\Phi(-\mu/\tau)},
\end{align*}
and the Bayes factors for other hypothesis comparisons can be obtained by
transitivity and reciprocity (e.g., $\mathrm{BF}_{-0} = \mathrm{BF}_{+0} \, / \,
\mathrm{BF}_{+-}$). Posterior probabilities can now be obtained by plugging the
Bayes factors and prior odds into~\eqref{eq:posterior}.

Bayes factors and posterior probabilities can be monitored as data accumulate to
see how the evidence for the hypotheses changes. Moreover, they can be plugged
into~\eqref{eq:prand} to obtain the randomization probability~\eqref{eq:mixture}
with $K = 1$, $\Pr(H_{+} \mid y, H_1) = \Phi(\mu_*/\tau_*)$, and
$\mathrm{BF}_{01}(\hat{\theta}) =
\exp[-\{(\hat{\theta}/\sigma_n)^2 - (\hat{\theta} - \mu)^2/(\sigma^2_n + \tau^2)
  \}/2] \sqrt{1 + \tau^2/\sigma^2_n}$.
As expected, the randomization probability shrinks towards equal randomization
as the prior probability of $H_0$ increases (i.e., $\pi \to 50\%$ as $\Pr(H_0)
\nearrow 1$), whereas it approaches the ordinary Bayesian posterior tail
probability of $\theta > 0$ based on a $\theta \sim \mathrm{N}(\mu, \tau^2)$
prior as the prior probability of $H_0$ decreases (i.e., $\pi \to 1 - \Phi\{(0 -
\mu_{*})/\tau_{*}\} = \Phi(\mu_*/\tau_*)$ as $\Pr(H_0) \searrow 0$). Similarly,
the randomization probability shrinks towards 50\% as the data provide more
evidence for $H_0$ ($\pi \to 50\%$ as $\mathrm{BF}_{01} \nearrow \infty$),
whereas it approaches the ordinary posterior tail probability as the evidence
for $H_1$ increases ($\pi \to \Phi(\mu_*/\tau_*)$ as $\mathrm{BF}_{01} \searrow
0$). In contrast to ad hoc modifications of Thompson sampling, shrinkage of the
posterior tail probability thus depends directly on the accumulated data via the
Bayes factor.
\end{sloppypar}

\begin{figure}[p]

\begin{knitrout}
\definecolor{shadecolor}{rgb}{0.969, 0.969, 0.969}\color{fgcolor}

{\centering \includegraphics[width=\maxwidth]{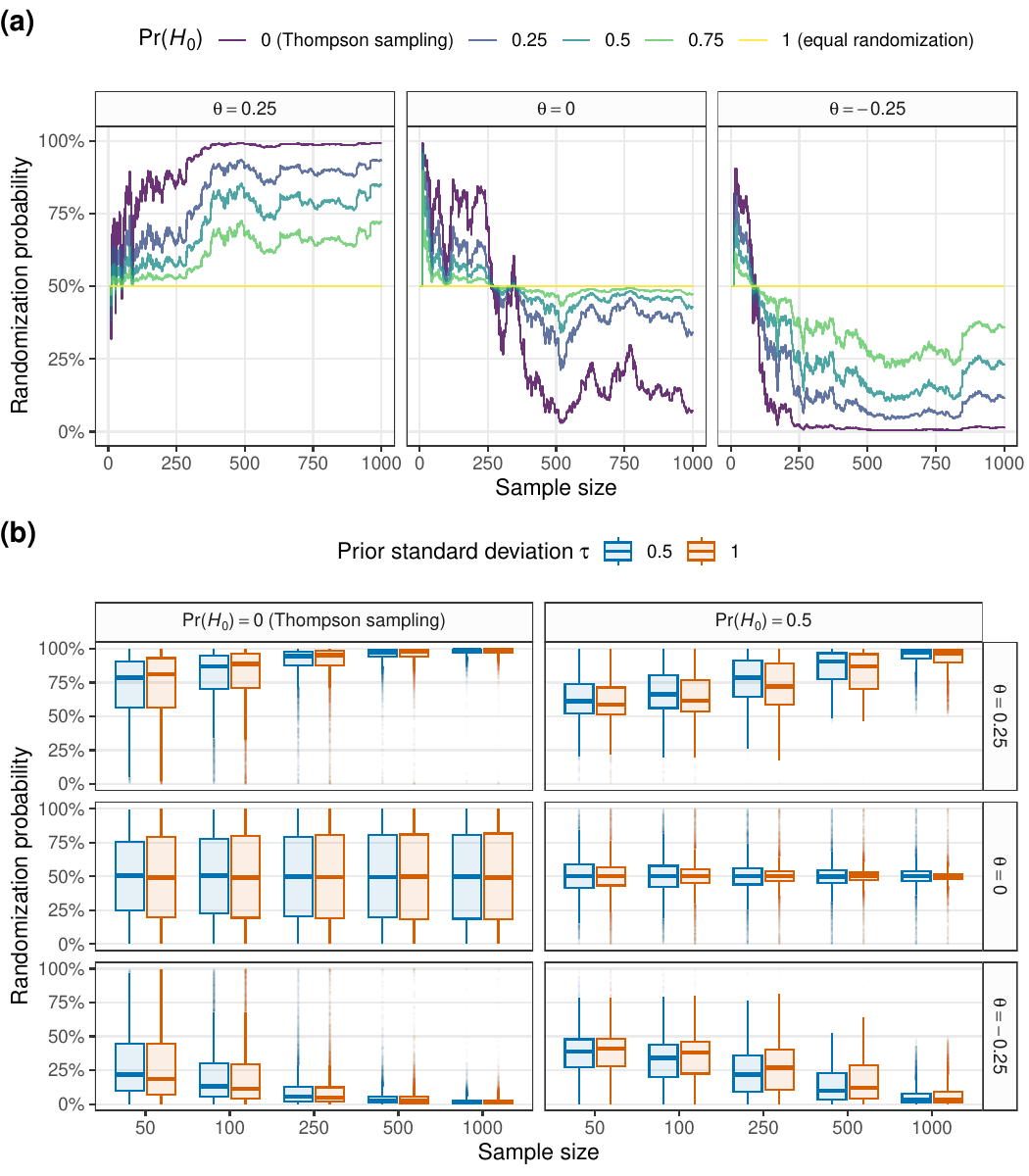} 

}

\end{knitrout}

\caption{Plot (a) shows the evolution of Bayesian RAR probabilities for three
  simulated data sequences. In each step, one patient is allocated to the
  treatment or control group using Thompson sampling assuming a normal prior
  centered at zero with standard deviation $\tau = 1$. Patient
  outcomes are then simulated from a normal distribution with a true standard
  deviation of 1 and assuming a true mean difference $\theta$ between
  treatment and control group as indicated in the plot panels. Randomization
  probabilities under the spike-and-slab prior ($\Pr(H_0) > 0$) are computed
  from the same data sequence to enable comparability. Plot (b) shows the
  distribution of randomization probabilities based on 10'000 simulations under Bayesian RAR for two different prior
  standard deviations and $\Pr(H_0)$.}
\label{fig:example-normal1}
\end{figure}

Figure~\ref{fig:example-normal1}a shows sequences of randomization probabilities
computed from normal data simulated under different true mean differences
between treatment and control group $\theta$ as indicated in the panels. For
each $\theta$, one sequence was simulated under RAR using Thompson sampling
($\Pr(H_0) = 0$). RAR probabilities for other values of $\Pr(H_0)$ were computed
from the same sequences to base comparison on the same data. Across all
$\theta$, the probability to randomize to the treatment group based on $\Pr(H_0)
= 1$ remains static at $50\%$ (yellow). In contrast, for a beneficial treatment
effect ($\theta > 0$; left plot), the randomization probabilities based on
$\Pr(H_0) < 1$ tend towards $100\%$ as more data accumulate, whereas for a
harmful treatment effect ($\theta < 0$; right plot), they tend towards 0\%.
Moreover, a clear ordering is visible: Probabilities under $\Pr(H_0) = 0$
(Thompson sampling) are the most extreme both in the ``correct'' and ``wrong''
direction. Probabilities based on $0 < \Pr(H_0) < 1$ show the same qualitative
behavior but are less extreme. Setting a higher $\Pr(H_0)$ thus reduces the
variability of randomization probabilities but also slows down convergence to
100\% or 0\%. Finally, when there is no treatment effect ($\theta = 0$; middle
plot), the probabilities based on $0 < \Pr(H_0) < 1$ stay relatively close to
50\%, whereas the Thompson sampling probabilities tend to wander around.

Figure~\ref{fig:example-normal1}b shows the distribution of randomization
probabilities obtained from 10'000 simulations of
the scenarios from Figure~\ref{fig:example-normal1}a and for two prior standard
deviations $\tau$. When the true mean difference is non-zero ($\theta = \pm
0.25$; first and third row), the randomization probabilities for
both $\Pr(H_0) = 0$ (Thompson sampling) and $\Pr(H_0) = 0.5$ correctly converge
to 100\% and 0\%, respectively, although the convergence of $\Pr(H_0) = 0.5$ is
slower. In contrast, when there is no difference ($\theta = 0$; second row),
only the randomization probabilities based on $\Pr(H_0) = 0.5$ converge to 50\%,
whereas the Thompson sampling randomization probabilities remain roughly
uniformly distributed for all sample sizes. The qualitative behavior of the
randomization probabilities is the same for both values of the prior standard
deviation $\tau$. However, convergence to 100\% and 0\% is slightly faster for
smaller $\tau$ (blue), whereas convergence to 50\% is faster for larger $\tau$
(orange).

\subsection{More than two groups}
\label{sec:multiplegroups}
Suppose now there are $K > 1$ treatment groups and consequently $K$ effect
estimates, each estimate quantifying the effect of the corresponding treatment
relative to control. A natural generalization is to stack them into a vector
$\boldsymbol{\hat{\theta}} = (\hat{\theta}_1, \dots, \hat{\theta}_K)^\top$ and
assume a $K$-variate normal distribution $\boldsymbol{\hat{\theta}} \,\vert\,
\boldsymbol{\theta} \sim \mathrm{N}_K(\boldsymbol{\theta},
\boldsymbol{\boldsymbol{\Sigma}}_n)$, where $\boldsymbol{\theta} = (\theta_1,
\dots, \theta_K)^\top$ is the vector of true effects and
$\boldsymbol{\boldsymbol{\Sigma}}_n$ is the (assumed to be known) covariance
matrix of $\boldsymbol{\hat{\theta}}$ that depends on the sample size $n$. For
example, $\boldsymbol{\hat{\theta}}$ could be a vector of estimated regression
coefficients and $\boldsymbol{\boldsymbol{\Sigma}}_n$ its estimated covariance
matrix.

The hypotheses from Section~\ref{sec:multiple-treatments} then translate into
$H_{-} \colon \max\{\theta_1, \dots, \theta_K\} < 0$ vs. $H_{0} \colon \theta_1
= \dots = \theta_K = 0$ vs. $H_{+i} \colon \theta_i = \max\{\theta_1, \dots,
\theta_K\} > 0$ for $i = 1, \dots, K$. All hypotheses apart from $H_0$ are
composite and require the specification of a prior distribution. In analogy to
the $K=1$ case, we specify a $K$-variate normal prior $\boldsymbol{\theta} \sim
\mathrm{N}_K(\boldsymbol{\mu}, \boldsymbol{\mathcal{T}})$ and truncate its
support to the region of the corresponding hypothesis. Similarly, a sensible
default is to center the prior on $\boldsymbol{0}$ to encode clinical equipoise,
and specify prior probabilities $\Pr(H_{+i})$ for $i=1, \dots, K$ so that the
$\mathrm{N}_K(\boldsymbol{\mu}, \boldsymbol{\mathcal{T}})$ distribution is
recovered when the prior is averaged over the hypotheses.

\begin{figure}[!htb]
\begin{knitrout}
\definecolor{shadecolor}{rgb}{0.969, 0.969, 0.969}\color{fgcolor}

{\centering \includegraphics[width=\maxwidth]{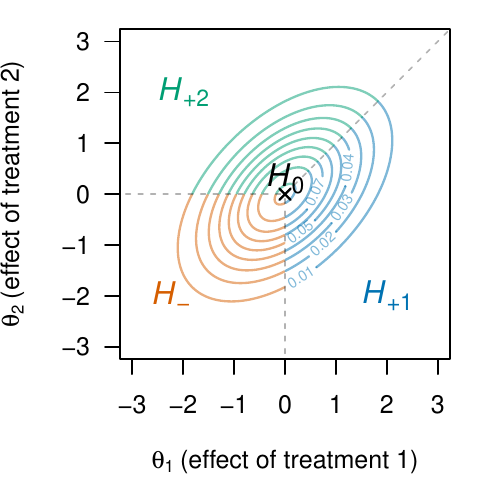} 

}

\end{knitrout}
\caption{Illustration of a spike-and-slab prior for a two-dimensional effect
  $\boldsymbol{\theta} = (\theta_1, \theta_2)^\top$. A point prior at $(0,
  0)^\top$ is assumed under $H_0$. A normal prior $\theta \sim
  \mathrm{N}((0, 0)^\top, \boldsymbol{\mathcal{T}})$ with
  $\boldsymbol{\mathcal{T}}_{ij} = 0.5$ for $i\neq j$ and
  $\boldsymbol{\mathcal{T}}_{ij} = 1$ for $i = j$, with support
  truncated to the space of the corresponding hypothesis is assumed under
  $H_{-}$, $H_{+1}$, and $H_{+2}$. The correlation of 0.5 ensures that all
  treatments receive equal prior probability $\Pr(H_{-}) = \Pr(H_{+1}) =
  \Pr(H_{+2}) = \{1 - \Pr(H_0)\}/3$.}
\label{fig:two-treatments-prior}
\end{figure}

Figure~\ref{fig:two-treatments-prior} illustrates a spike-and-slab prior for two
treatment groups ($K = 2$). The prior covariance matrix
$\boldsymbol{\mathcal{T}}$ is set to a uniform correlation of 0.5 which ensures
that the prior probabilities of all hypotheses but $H_0$ are equal. Note that a
zero-centered prior with uniform correlation of 0.5 ensures equal treatment
prior probabilities for any number of treatment groups $K$, and thus is a
sensible default in the absence of prior knowledge.

Also for $K > 1$, marginal likelihoods, Bayes factors, posterior probabilities,
and randomization probabilities are available in closed form (shown in Section~B
of the supplement). Crucially, only the evaluation of multivariate normal
densities and CDFs is required. This thus leads to an efficient Bayesian RAR
method that allows experimenters to compute randomization probabilities in
complex settings, for instance, multiple regression, where a full Bayesian
analysis may involve additional complexities, such as priors for nuisance
parameters and Markov chain Monte Carlo methods for the computation of
posteriors.

\section{Null hypothesis Bayesian RAR for binary outcomes}
\label{sec:binary}
Suppose now that we observe binary data of the form $y = \{y_C, y_1, \dots,
y_{K}\}$ and $n = \{n_C, n_1, \dots, n_{K}\}$ where $y_i$ denotes the number of
successes out of $n_i$ patients in group $i \in \{C, 1, \dots, K\}$ coming from a
control group ($C$) and $K$ treatment groups. All success counts are assumed to
be binomially distributed with probabilities $\theta_C, \theta_1, \dots,
\theta_{K}$, respectively, with higher values indicating greater benefit. The
hypotheses from Section~\ref{sec:multiple-treatments} then translate into $H_{-}
\colon \theta_C = \max\{\theta_C, \theta_1, \dots, \theta_K\}$ vs. $H_0 \colon
\theta_C = \theta_1 = \dots = \theta_{K}$ vs. $H_{+i} \colon \theta_i =
\max\{\theta_C, \theta_1, \dots, \theta_K\}$ for $i = 1, \dots, K$. In the
normal framework from Section~\ref{sec:normal}, these could be translated into
hypotheses related to log odds ratios, which can be estimated with logistic
regression or other methods. However, normal approximations can be inaccurate
for small sample sizes or extreme probabilities, so exact binomial computation
is preferable.

The null hypothesis $H_0$ requires specification of a prior for the common
probability $\theta_C$. Assuming a beta prior $\theta_C \,\vert\, H_0 \sim
\mathrm{Beta}(a_0, b_0)$, the marginal likelihood under $H_0$ is
\begin{align*}
  \Pr(y \,\vert\, H_0, n)
  =& \prod_{j\in\{C,1,\dots,K\}} \binom{n_j}{y_j} \times \frac{\mathrm{B}(a_0 + \sum_{i\in\{C,1,\dots,K\}} y_i, b_0 + \sum_{i\in\{C,1,\dots,K\}} n_i - y_i)}{\mathrm{B}(a_0, b_0)}
\end{align*}
with beta function $\mathrm{B}(\cdot, \cdot)$. Under the other hypotheses, it is
natural to assign independent beta priors $\theta_i \sim \mathrm{Beta}(a_{i},
b_i)$ for $i \in \{C, 1, \dots, K\}$, and truncate their support to the space of
the corresponding hypothesis. This leads to the marginal likelihood of the data
\begin{align*}
  \Pr(y \,\vert\, H_i, n)
  &= \prod_{j\in\{C,1,\dots,K\}} \binom{n_j}{y_j} \times \frac{\mathrm{B}(a_j + y_j, b_j + n_j - y_j)}{\mathrm{B}(a_j, b_j)} \\
  & \times \frac{Q_i(a_C + y_C, a_1 + y_1, \dots, a_{K} + y_{K}, b_C + n_C - y_C, b_1 + n_1 - y_1, \dots, b_{K} + n_{K} - y_{K})}{Q_i(a_C, a_1, \dots, a_{K}, b_C, b_1, \dots, b_{K})}
\end{align*}
under $H_i \in \{H_{-}, H_{+1}, \dots, H_{+K}\}$ and with
\begin{align*}
  Q_i(a_C, a_1, \dots, a_{K}, b_C, b_1, \dots, b_{K})
  &= \int_0^1 \frac{\theta_i^{a_i - 1} (1 - \theta_i)^{b_i - 1}}{\mathrm{B}(a_i, b_i)} \times \prod_{j\in\{C,1,\dots,K\} \setminus \{i\}} I_{\theta_i}(a_j, b_j) \, \mathrm{d}\theta_i,
\end{align*}
where $i = C$ under $H_{-}$ and $i$ is the index of the superior treatment under
$H_{+i}$, and where $I_{x}(a, b)$ is
the CDF of the $\mathrm{Beta}(a, b)$ distribution evaluated at $x$. These
marginal likelihoods can be efficiently computed with numerical integration. For
priors with integer hyperparameters, there even exist closed-form solutions
\citep{Howard1998, Miller2015}.

For a specified $\Pr(H_0)$, we may again distribute the remaining prior
probability among the other hypotheses by $\Pr(H_{+i}) = \{1 - \Pr(H_0)\} \times
Q_i(a_C, a_1, \dots, a_{K}, b_C, b_1, \dots, b_{K})$ to ensure that the averaged
prior is again the beta prior. Specifying uniform priors ($a_j = b_j = 1 ~
\forall ~ j \in \{C, 1, \dots, K\}$) is an intuitive default that ensures that
each treatment receives equal prior probability. Plugging the marginal
likelihoods and prior probabilities into~\eqref{eq:posterior} produces posterior
probabilities, which in turn can be used to obtain randomization probabilities.

\section{Reanalyzing the ECMO trial}
\label{sec:applications}
The ECMO trial \citep{Bartlett1985, Bartlett2024} investigated ECMO
(extracorporeal membrane oxygenation) treatment in critically ill newborns.
While earlier non-randomized studies suggested a substantial treatment effect, a
randomized study was required to confirm this. However, investigators were
convinced the mortality risk was much higher in the control group, posing an
ethical dilemma. To mitigate this, a ``randomized play-the-winner'' (RPW) RAR
design \citep{Wei1978} was chosen.

The trial's outcome was extreme: The first newborn was randomized to ECMO and
survived. The second was randomized to the control and died. All ten
subsequently enrolled newborns were randomized to ECMO and survived. The trial
was then stopped for efficacy, yet its unusual outcome sparked debate about
whether it constituted a proper randomized trial. Later trials confirmed the
efficacy of ECMO, which is now a standard treatment.

\begin{figure}[!tb]
\begin{knitrout}
\definecolor{shadecolor}{rgb}{0.969, 0.969, 0.969}\color{fgcolor}

{\centering \includegraphics[width=\maxwidth]{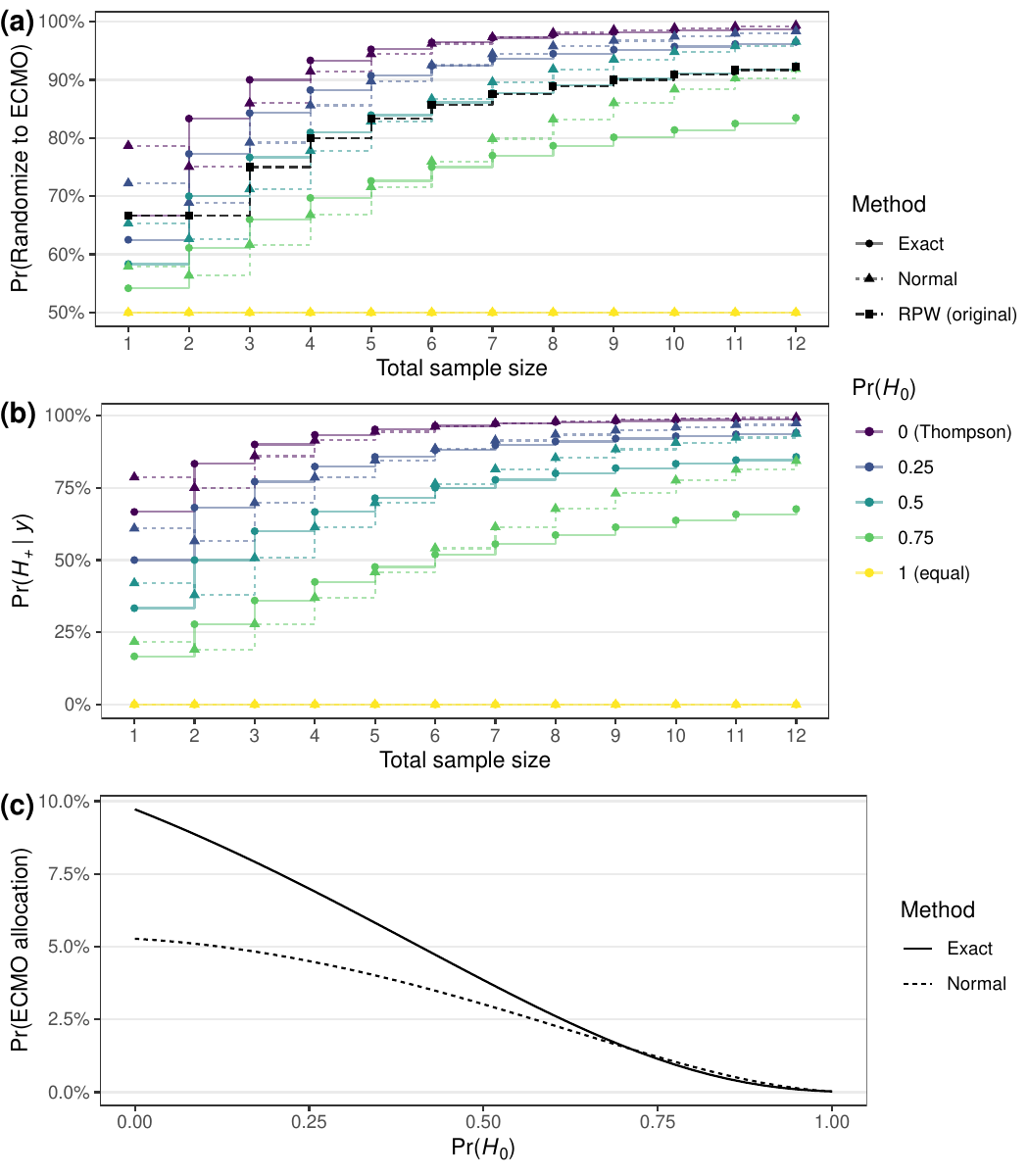} 

}

\end{knitrout}
\caption{Evolution of Bayesian RAR randomization probabilities (a) and posterior
  probability of a beneficial ECMO treatment effect (b) for data from the ECMO
  trial. Plot (c) shows the probability of observing the ECMO allocation
  sequence as a function of $\Pr(H_0)$.
}
\label{fig:ECMO}
\end{figure}

Figure~\ref{fig:ECMO}a shows RAR probabilities computed from the ECMO data. The
normal approximation and exact binomial methods were used, as well as the RPW
method originally used in the ECMO trial. A standard normal prior was assigned
to the log odds ratio for the normal method, while uniform priors were assigned
to the probabilities for the binomial method. Log odds ratio estimates were
computed by adding a half to each cell to avoid zero cell issues. This
correction induces a slight anomaly, as the probabilities from the normal method
slightly decrease after the first newborn even though it survived. This does
not happen for the exact method whose probabilities increase after every
additional patient. Furthermore, the randomization probabilities differ notably
between the normal and exact methods, even at the end of the study. This
presumably happens because the influence of the differing priors remains
relatively high after observing only 12 patients.

Comparing the randomization probabilities for different prior probabilities of
the null hypothesis, $\Pr(H_0) = 0$ (Thompson sampling) shows the most extreme
randomization probabilities that rapidly increase to 100\%, whereas $\Pr(H_0) =
1$ (equal randomization) leaves the probabilities completely static at 50\%. In
between, the probabilities gradually shift from 50\% to the Thompson sampling
probabilities. The RPW probabilities (black squares) are relatively close to the
exact method with $\Pr(H_0) = 0.5$ at the beginning of the study and become
closer to the normal method with $\Pr(H_0) = 0.75$ at later times.

Figure~\ref{fig:ECMO}b shows the corresponding posterior probability of the ECMO
treatment being effective. Depending on $\Pr(H_0)$, the posterior probability at
the end of the trial may be very high (e.g., $\Pr(H_{+} \,\vert\, y) =
0.99$ for $\Pr(H_0) = 0$), or only moderately favoring ECMO
(e.g., $\Pr(H_{+} \,\vert\, y) = 0.86$ for $\Pr(H_0) = 0.5$).
Stopping the trial thus seems sensible only if the prior probability of no ECMO
effect was low.

Figure~\ref{fig:ECMO}c shows the probability of observing the ECMO allocation
sequence under Bayesian RAR with different $\Pr(H_0)$. The probability is
highest for $\Pr(H_0) = 0$ (Thompson sampling) and decreases with increasing
$\Pr(H_0)$ to $(1/2)^{12} = 0.024\%$ for $\Pr(H_0) =
1$ (equal randomization). Interestingly, the probability under RPW is also
rather low (2.8\%). Hence, observing the ECMO allocation
sequence under Bayesian RAR would only be less likely than under RPW for
$\Pr(H_0) > 0.6$ (exact method) and $\Pr(H_0) >
0.55$ (normal method), respectively.

\section{Simulation study}
\label{sec:simulation}
We conducted a simulation study to evaluate the performance of null hypothesis
Bayesian RAR for different values of $\Pr(H_0)$, and compare it to Thompson
sampling (potentially modified with burn-in periods, probability capping, or power
transformations), equal randomization, the Gittins index \citep{Gittins2011},
and Bayes upper confidence bound (UCB) \citep{Kaufmann2012}. Considered patient
benefit measures were the mean rate of successes, the rate of extreme
randomization probabilities, and sample size imbalance in favor of the inferior
treatment. Bias and coverage were used to evaluate performance of response rate
difference point estimates and confidence intervals under RAR, while the type I
error rate and power of the corresponding tests were used to quantify hypothesis
testing performance. A binomial data-generating mechanism was used. The response
probability in the control group was fixed to $\theta_C = 0.25$ and varied in
the first treatment group $\theta_1 \in \{0.25, 0.35, 0.45\}$ to assess the
behavior of the methods for different effect sizes $\text{RD}_1 = \theta_1 -
\theta_C$. In conditions with two or three treatment groups, the corresponding
probabilities were set to $\theta_2 = \theta_3 = 0.3$. More detailed
descriptions of the design and results of the simulation study
are provided in Section~C of the supplement.

\begin{figure}[p]
\begin{knitrout}
\definecolor{shadecolor}{rgb}{0.969, 0.969, 0.969}\color{fgcolor}

{\centering \includegraphics[width=\maxwidth]{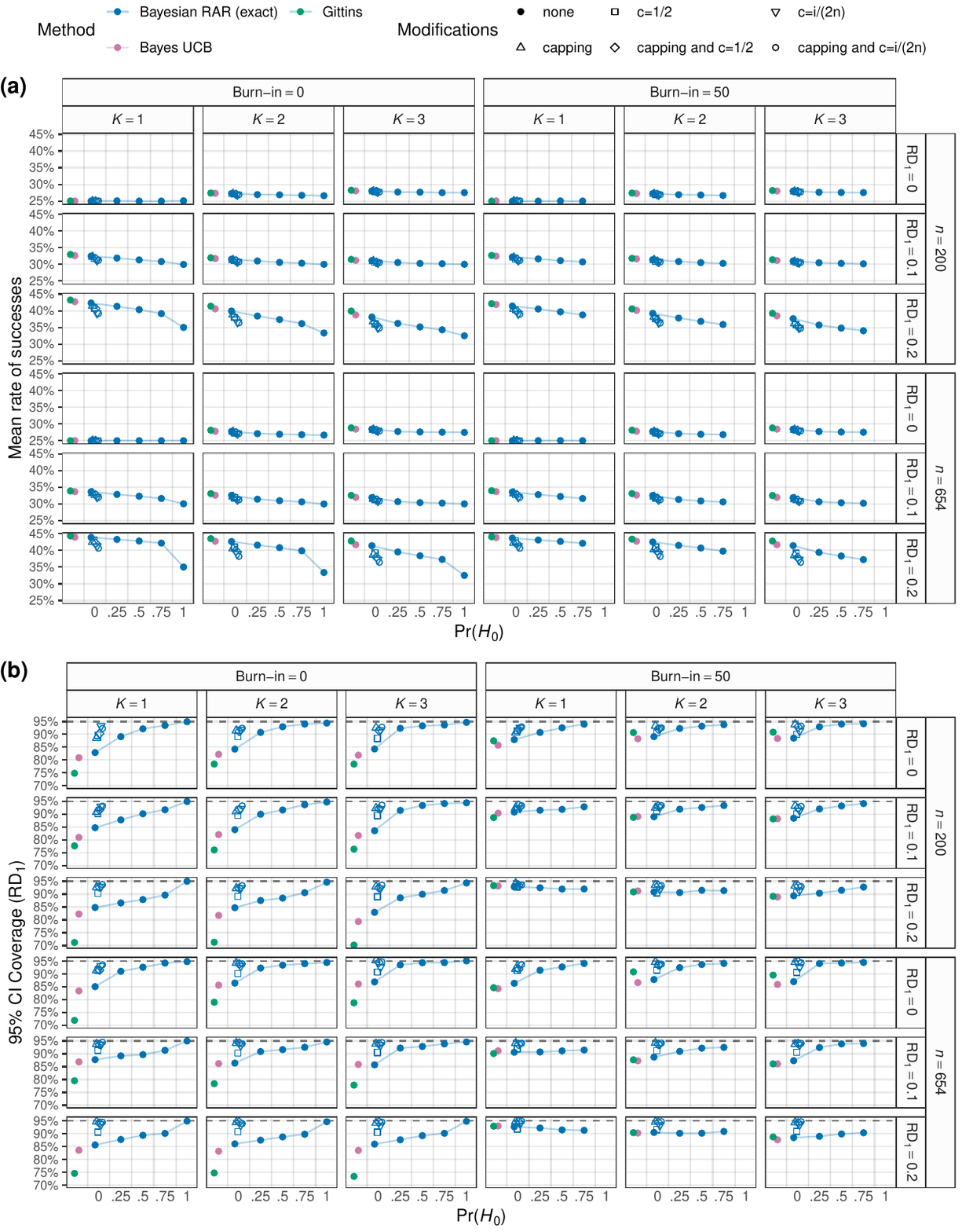} 

}

\end{knitrout}
\caption{Plot (a) shows the mean rate of successes (i.e., the number of
  successes in a study divided by its sample size averaged over all
  10'000 simulation repetitions). The maximum
  Monte Carlo standard error (MCSE) is
  0.051\%. Plot (b) shows the
  empirical coverage rate of the 95\% Wald confidence interval for the response
  rate difference $\text{RD}_1$. The maximum MCSE is
  0.46\%. Gittins index and
  Bayes UCB do not depend on $\Pr(H_0)$ and are shown to the left of $\Pr(H_0) =
  0$ for comparison.}
\label{fig:simulationmain}
\end{figure}

A trade-off between patient benefit and inferential operating characteristics
was observed across most settings. Methods that improved patient benefit through
more aggressive allocation (Gittins index, Bayes UCB, and Thompson sampling)
often had worse bias, coverage, type I error rate, and power, whereas more
conservative methods tended to preserve inferential validity at the cost of
reduced patient benefit. Figure~\ref{fig:simulationmain} illustrates this
trade-off for the mean success rate and coverage. This trade-off is well
documented \citep{Hu2003, Zhang2005, Williamson2019, Robertson2023}. However,
there were also settings with a small ``window'' in which patient benefit could
vary while inferential performance remained essentially unchanged. For example,
with a burn-in of 50, two treatment groups ($K = 2$), and a large effect size
($\text{RD}_1 = 0.2$), coverage was roughly constant across $\Pr(H_0)$, whereas
the mean success rate changed noticeably. Increasing $\Pr(H_0)$ reduced
imbalance toward the inferior treatment, though negative imbalance was also
strongly influenced by effect size. In conditions with large effect size
($\text{RD}_1 = 0.2$), the proportion of negative imbalance was often lower for
RAR methods (including Thompson sampling) than for equal randomization, thus
confirming the results from \citet{Robertson2023}. These findings highlight the
importance of evaluating operating characteristics during trial design, since
appropriately specified RAR designs may enable improved patient benefit without
substantial loss of inferential performance.

Under most conditions, Bayesian RAR with $\Pr(H_0) = 0.75$ showed similar
operating characteristics to Thompson sampling with capped randomization
probabilities at 10\%/90\% and power transformation $c=i/(2n)$, where $i$ is the
current sample size and $n$ is the maximum sample size. Both mitigated issues
with ordinary Thompson sampling, exhibiting less negative sample size imbalance,
less bias, improved coverage, and reduced type I error rates. This came at the
cost of worse patient benefit performance, for instance, lower mean success
rate, though still better than equal randomization in most cases. Smaller values
of $\Pr(H_0)$ produced operating characteristics comparable to less extreme
modifications, such as $c = 1/2$. Finally, the results were similar for exact
binomial and approximate normal versions of the method (Section~C of the
supplement), suggesting some robustness to prior and model choice.

While these results demonstrated that null hypothesis Bayesian RAR has
comparable properties to capped Thompson sampling for realistic trial sizes, it
is important to note that the two are expected to differ asymptotically. Capped
Thompson sampling enforces bounds on randomization probabilities, whereas
uncapped Bayesian RAR probabilities can converge to 100\% or 0\%. The observed
similarity is thus only a finite-sample phenomenon.

\section{Discussion}
\label{sec:discussion}

We have proposed a modification of Thompson sampling by recasting the problem in
a Bayesian hypothesis testing framework and introducing a null hypothesis of
equal effectiveness. The method can interpolate between equal randomization and
Thompson sampling by changing the prior probability of the null hypothesis
$\Pr(H_0)$. This allows balancing patient benefit with inferential performance.
For large values of $\Pr(H_0)$, we observed similar characteristics to those of ad
hoc modifications of Thompson sampling.

One advantage of the method is that randomization probabilities coherently
correspond to the available statistical evidence (in the form of Bayes factors)
and beliefs (in the form of posterior probabilities). Both could also serve as
decision-making tools instead of frequentist criteria. For example, a study
could be stopped if the posterior probability of efficacy exceeds 0.99. This
also aligns with the perspective that it is unnatural to randomize patients
using extreme randomization probabilities associated with such high posterior
probabilities.

Although we conducted a simulation study to understand the method's basic
behavior, more evaluations are needed to understand its real-world
applicability. For example, the method needs to be evaluated in combination with
futility stopping (e.g., dropping of arms at interim) and under time trends
(e.g., effects attenuating over time because the standard of care improves). RAR
is often impractical when primary outcomes require long follow-up. An
alternative could be to perform Bayesian RAR with a surrogate outcome that is
observed sooner \citep{Gao2024}. Finally, further theoretical work is needed to
better understand the asymptotic behavior of the method under non-normal models
such as the binomial model considered in Section~\ref{sec:binary}.

Regarding prior choice, $\Pr(H_{0}) = 0.75$ mitigated many Thompson sampling
issues in our study, yet higher values may also be considered. Similar
considerations apply to the prior distributions of the parameters. For instance,
rather than setting the correlation of the multivariate normal prior to achieve
equal randomization probabilities, it could be set so that the prior probability
of the control being superior is always $\Pr(H_{-}) = 0.5$ and the remaining
probability is equally distributed among the treatments. Other choices of the
correlation structure may also improve operating characteristics. Similarly,
concurrent allocation to the control group may be desirable \citep{Trippa2012}.
Future work may therefore investigate whether more efficient RAR can be obtained
by tuning the prior distribution.

\paragraph{Acknowledgments and conflict of interest}
We thank František Bartoš, Małgorzata Roos, and Manuela Ott for valuable
comments. We thank the associate editor and two anonymous reviewers for
excellent suggestions, and especially one reviewer for providing the proof
sketch establishing that the randomization probabilities converge to equal
randomization under $H_0$. We declare no conflict of interest.

\paragraph{Data availability}
Code and data to reproduce our results as well as the R package \texttt{brar}
are available at \url{https://github.com/SamCH93/brar} and archived at
\url{https://doi.org/10.5281/zenodo.17248628}. The R package is also available
on CRAN (\url{https://CRAN.R-project.org/package=brar}). Section~D in the
supplement illustrates usage of the package.

\bibliographystyle{apalikedoiurl}
\bibliography{bibliography}

\begin{appendices}

\counterwithin{figure}{section}
\counterwithin{table}{section}
\renewcommand{\thesection}{\Alph{section}}
\renewcommand{\thefigure}{\thesection\arabic{figure}}
\renewcommand{\thetable}{\thesection\arabic{table}}

\section{Asymptotic randomization probabilities under the null}
\label{sec:supasymptotic}

Define the alternative hypothesis $H_1$ as the complement of $H_0$ or
equivalently as the union of all non-null hypotheses, i.e., $H_1 =
\bigcup_{j\neq 0} H_j = H_{-} \cup H_{+1} \cup \dots \cup H_{+K}$. The
probability to randomize to treatment group $i$ can then be expressed as
\begin{align*}
  \pi_i
  &= \Pr(H_{+i} \mid y) + \Pr(H_0 \mid y)/(K + 1) \\
  &= \Pr(H_{+i} \mid y, H_1) \times \Pr(H_{1} \mid y) + \{1 - \Pr(H_1 \mid y)\}/(K + 1) \\
  &= w \times \underbrace{\Pr(H_{+i} \mid y, H_1)}_{\text{Thompson sampling}} +
  (1 - w) \times \underbrace{1/(K + 1)}_{\text{equal randomization}}
\end{align*}
with weight
\begin{align*}
  w &= 1 \, \bigg / \, \left\{1 + \frac{\Pr(H_0)}{1 - \Pr(H_0)} \times \mathrm{BF}_{01}(y)\right\}.
\end{align*}
If the Bayes factor $\mathrm{BF}_{01}(y)$ goes to zero, the weight converges to
one and consequently the randomization probability converges to the ordinary
Thompson sampling randomization probability (i.e., the posterior probability of
$H_{+i}$ conditional on $H_1$). Conversely, if the Bayes factor
$\mathrm{BF}_{01}(y)$ diverges to $+\infty$, the weight goes to zero and
consequently the randomization probability converges to equal randomization.

Denote by $Y_n$ the random data with sample size $n$. The inverse Bayes factor
$1/\mathrm{BF}_{01}(Y_n)$ is a test martingale for $H_0$ \citep{Shafer2011}.
This also holds under RAR because the allocation is predictable (the
randomization probability for the $n$-th patient is computed from the previously
observed data $Y_{n-1}$ and is thus fixed before the outcome of patient $n$ is
observed). It then follows from Ville's inequality that
\begin{align*}
  \Pr\left\{\sup_{n > 0} 1/\mathrm{BF}_{01}(Y_n) \geq 1/\alpha \mid H_0\right\} =
  \Pr\left\{\inf_{n > 0} \mathrm{BF}_{01}(Y_n) \leq \alpha \mid H_0\right\}
  \leq \alpha
\end{align*}
for every $\alpha \leq 1$. Letting $\alpha \to 0$ shows that, under $H_0$, the
Bayes factor is almost surely bounded away from zero. Consequently, the
equal-randomization weight $1 - w$ is almost surely bounded away from zero for
all $n$, so the control group and every treatment group have allocation
probabilities that are bounded away from zero uniformly in $n$. Hence, every
group is sampled infinitely often almost surely.
By Theorem 3.1 from \citet{Melfi2000}, ordinarily consistent estimators thus
remain consistent under null hypothesis Bayesian RAR if $H_0$ is true.

To show that the Bayes factor diverges to $+\infty$ and randomization
probabilities consequently tend to equal randomization probabilities, further
assumptions on the data and form of the Bayes factor are required. In the
following, we will assume the setting of Section~3 of the main text. We start
with the one treatment group case ($K=1$). Suppose the Bayes factor concerns the
test of a mean difference $\theta$ based on the maximum likelihood estimate
$\hat{\theta}_n$ with standard error $\sigma_n = \lambda \sqrt{1/n_C + 1/n_1}$.
The logarithm of the Bayes factor is given by
\begin{align}
  \label{eq:logbf01}
  \log \mathrm{BF}_{01} = \frac{1}{2}\left\{\frac{(\hat{\theta}_n -
    \mu)^2}{\sigma^2_n + \tau^2} + \log \left(1 +
  \frac{\tau^2}{\sigma^2_n}\right) - \frac{\hat{\theta}_n^2}{\sigma^2_n}
  \right\}.
\end{align}
Since under $H_0 \colon \theta = 0$, each arm is sampled infinitely often, it
follows that $\sigma_n$ tends to $0$ and that $\hat{\theta}_n$ converges to the
true mean difference $\theta$. With decreasing $\sigma_n$, the first term
in~\eqref{eq:logbf01} approaches the constant $\mu^2/\tau^2$ and the second term
diverges to $+\infty$. The last term $-(\hat{\theta}_n/\sigma_n)^2$ involves the
squared standardized estimate, whose growth we now bound.

Conditionally on the realized sample sizes $n_C$ and $n_1$, and assuming that
the standard deviation $\lambda$ is known, the expectation of $\hat{\theta}_n$
equals 0 under $H_0$ and its variance equals $\sigma_n^2$, so that
\begin{align*}
  \mathbb{E}[(\hat{\theta}_n/\sigma_n)^2 \mid H_0, n_C, n_1] =
  \operatorname{Var}(\hat{\theta}_n/\sigma_n \mid H_0, n_C, n_1) = 1.
\end{align*}
Thus, by Markov's inequality,
\begin{align*}
  \Pr\{(\hat{\theta}_n/\sigma_n)^2 \geq 1/\alpha \mid H_0, n_C, n_1\} \leq \frac{\mathbb{E}[(\hat{\theta}_n/\sigma_n)^2 \mid H_0, n_C, n_1]}{1/\alpha} = \alpha
\end{align*}
for every $\alpha \leq 1$. Averaging over the (adaptively generated)
distribution of $n_C$ and $n_1$ leaves the bound intact
\begin{align*}
  \Pr\{(\hat{\theta}_n/\sigma_n)^2 \geq 1/\alpha \mid H_0\} =
  \mathbb{E}_{n_C, n_1}\left[\Pr\{(\hat{\theta}_n/\sigma_n)^2 \geq 1/\alpha \mid H_0, n_C, n_1\}\right]
  \leq \alpha,
\end{align*}
as the bound does not depend on $n_C$ and $n_1$. Hence,
$(\hat{\theta}_n/\sigma_n)^2$ is bounded in probability under $H_0$. Therefore,
since the diverging term $\log(1 + \tau^2/\sigma_n^2)$ in~\eqref{eq:logbf01}
dominates the stochastically bounded term $(\hat{\theta}_n/\sigma_n)^2$, the log
Bayes factor diverges to $+\infty$ in probability. Consequently, under $H_0$ the
allocation probabilities converge to equal randomization in probability.

Using a similar argument, this can also be shown for the case with more than one
treatment group ($K > 1$). Here, the logarithm of the Bayes factor is
\begin{align}
  \label{eq:logbf01multi}
  \log \mathrm{BF}_{01} =
  \frac{1}{2} \left\{(\boldsymbol{\hat{\theta}}_n - \boldsymbol{\mu})^\top (\boldsymbol{\Sigma}_n + \boldsymbol{\mathcal{T}})^{-1} (\boldsymbol{\hat{\theta}}_n - \boldsymbol{\mu}) +
  \log\left(\frac{|\boldsymbol{\Sigma}_n + \boldsymbol{\mathcal{T}}|}{|\boldsymbol{\Sigma}_n|} \right)
  - \boldsymbol{\hat{\theta}}^\top_n \boldsymbol{\Sigma}^{-1}_n \boldsymbol{\hat{\theta}}_n \right\}.
\end{align}
As under $H_0$ every group is sampled infinitely often,
$\boldsymbol{\hat{\theta}}_n \to \boldsymbol{0}$, $\boldsymbol{\Sigma}_n \to
\boldsymbol{0}$ (a matrix of zeros), and $|\boldsymbol{\Sigma}_n| \rightarrow
0$. The first term in~\eqref{eq:logbf01multi} thus converges to the constant
$\boldsymbol{\mu}^\top\boldsymbol{\mathcal{T}}^{-1}\boldsymbol{\mu}$, while the
second term diverges to $+\infty$. The final quadratic form
$\boldsymbol{\hat{\theta}}^\top_n \boldsymbol{\Sigma}^{-1}_n
\boldsymbol{\hat{\theta}}_n$ has, conditional on $H_0$ and the realized sample
sizes $n_C, n_1, \dots, n_K$, expectation
\begin{align*}
  \mathbb{E}[\boldsymbol{\hat{\theta}}^\top_n \boldsymbol{\Sigma}^{-1}_n
\boldsymbol{\hat{\theta}}_n \mid H_0, n_C, n_1, \dots, n_K]
  &= \mathbb{E}[\operatorname{tr}(\boldsymbol{\hat{\theta}}^\top_n \boldsymbol{\Sigma}^{-1}_n
\boldsymbol{\hat{\theta}}_n) \mid H_0, n_C, n_1, \dots, n_K] \\
  &= \mathbb{E}[\operatorname{tr}(\boldsymbol{\Sigma}^{-1}_n
\boldsymbol{\hat{\theta}}_n \boldsymbol{\hat{\theta}}^\top_n) \mid H_0, n_C, n_1, \dots, n_K] \\
  &= \operatorname{tr}(\boldsymbol{\Sigma}^{-1}_n
\mathbb{E}[\boldsymbol{\hat{\theta}}_n \boldsymbol{\hat{\theta}}^\top_n \mid H_0, n_C, n_1, \dots, n_K]) \\
&= \operatorname{tr}\{\boldsymbol{\Sigma}^{-1}_n
\operatorname{Var}(\boldsymbol{\hat{\theta}}_n \mid H_0, n_C, n_1, \dots, n_K)\} \\
  & = \operatorname{tr}(\boldsymbol{\Sigma}^{-1}_n \boldsymbol{\Sigma}_n) \\
&= K.
\end{align*}
Thus, by Markov's inequality $\Pr(\boldsymbol{\hat{\theta}}^\top_n
\boldsymbol{\Sigma}^{-1}_n \boldsymbol{\hat{\theta}}_n \geq 1/\alpha \mid H_0,
n_C, n_1, \dots, n_K) \leq K \alpha$ for every $\alpha \leq 1/K$, a bound that
does not depend on the sample sizes. Averaging over their distribution leaves
the bound intact, so the quadratic form is bounded in probability. The diverging
log-determinant term in~\eqref{eq:logbf01multi} therefore dominates, and $\log
\mathrm{BF}_{01} \to +\infty$ in probability, so that under $H_0$ randomization
probabilities converge to equal randomization also for the case of multiple
treatment groups.

\section{Null hypothesis Bayesian RAR under multivariate normality}
Suppose there are $K > 1$ treatment groups and consequently $K$ effect
estimates, each estimate quantifying the effect of the corresponding treatment
relative to the control. A natural generalization of univariate normal Bayesian
RAR is to stack the estimates into a vector $\boldsymbol{\hat{\theta}} =
(\hat{\theta}_1, \dots, \hat{\theta}_K)^\top$ and assume a $K$-variate normal
distribution $\boldsymbol{\hat{\theta}} \mid \boldsymbol{\theta} \sim
\mathrm{N}_K(\boldsymbol{\theta}, \boldsymbol{\Sigma}_n)$, where
$\boldsymbol{\theta} = (\theta_1, \dots, \theta_K)^\top$ is the vector of true
effects and $\boldsymbol{\boldsymbol{\Sigma}}_n$ is the (assumed to be known)
covariance matrix of $\boldsymbol{\hat{\theta}}$ that depends on the sample size
$n$. For example, $\boldsymbol{\hat{\theta}}$ could be a vector of estimated
regression coefficients and $\boldsymbol{\Sigma}_n$ its estimated
covariance matrix.

The hypotheses from Section~2.1 in the main text then translate into
\begin{align*}
  &H_{-} \colon \max\{\theta_1, \dots, \theta_K\} < 0&
   &\text{vs.} &
  &H_{0} \colon \theta_1 = \dots = \theta_K = 0&
   &\text{vs.} &
  &H_{+i} \colon \theta_i = \max\{\theta_1, \dots, \theta_K\} > 0&
\end{align*}
for $i = 1, \dots, K$. All hypotheses apart from $H_0$ are composite and require
the specification of a prior distribution. In analogy to the one treatment case,
we specify a $K$-variate normal prior $\boldsymbol{\theta} \sim
\mathrm{N}_K(\boldsymbol{\mu}, \boldsymbol{\mathcal{T}})$ and truncate its
support to the region of the corresponding hypothesis (e.g., for hypothesis
$H_{+i}$, the space in $\mathbb{R}^K$ where the $i$th component is positive and
larger than all other components). As in the one group case, it seems a
sensible default to center the prior on $\boldsymbol{0}$ to encode clinical
equipoise. Similarly, the prior hypothesis probabilities $\Pr(H_{+i})$ for $i=1,
\dots, K$ may again be specified so that the $\mathrm{N}_K(\boldsymbol{\mu},
\boldsymbol{\mathcal{T}})$ distribution is recovered when the prior is averaged
over the hypotheses. Additionally, specifying a prior covariance matrix
$\boldsymbol{\mathcal{T}}$ with uniform correlation of 0.5 ensures equal
treatment prior probabilities for any number of treatment groups $K$, and thus
seems a sensible default in the absence of prior knowledge.

\begin{sloppypar}
The normal-normal conjugate framework allows us to derive marginal likelihoods
in closed form. The marginal likelihood under $H_{0}$ is
\begin{align*}
  p(\boldsymbol{\hat{\theta}} \mid H_{0}, n)
  &= \mathrm{N}_K(\boldsymbol{\hat{\theta}} \mid \boldsymbol{0}, \boldsymbol{\Sigma}_n),
\end{align*}
while the marginal likelihood under $H_{-}$ is
\begin{align*}
  p(\boldsymbol{\hat{\theta}} \mid H_{-}, n)
  &= \mathrm{N}_K(\boldsymbol{\hat{\theta}} \mid \boldsymbol{\mu}, \boldsymbol{\Sigma}_n + \boldsymbol{\mathcal{T}}) \times \frac{\Phi_K(\boldsymbol{0} \mid \boldsymbol{\mu}_*, \boldsymbol{\mathcal{T}}_*)}{\Phi_K(\boldsymbol{0} \mid \boldsymbol{\mu}, \boldsymbol{\mathcal{T}})}
\end{align*}
with $\mathrm{N}_K(\boldsymbol{x} \mid \boldsymbol{m}, \boldsymbol{V})$ and
$\Phi_K(\boldsymbol{x} \mid \boldsymbol{m}, \boldsymbol{V})$ the density and
cumulative distribution functions of the $K$-variate normal distribution with
mean vector $\boldsymbol{m}$ and covariance matrix $\boldsymbol{V}$ evaluated at
$\boldsymbol{x}$, and posterior mean $\boldsymbol{\mu}_* =
(\boldsymbol{\Sigma}^{-1}_n +
\boldsymbol{\mathcal{T}}^{-1})^{-1}(\boldsymbol{\Sigma}^{-1}_n\boldsymbol{\hat{\theta}}
+ \boldsymbol{\mathcal{T}}^{-1}\boldsymbol{\mu})$ and covariance
$\boldsymbol{\mathcal{T}}_* = (\boldsymbol{\Sigma}^{-1}_n +
\boldsymbol{\mathcal{T}}^{-1})^{-1}$. Finally, the marginal likelihood under
$H_{+i}$ is given by
\begin{align*}
  p(\boldsymbol{\hat{\theta}} \mid H_{+i}, n)
  &= \mathrm{N}_K(\boldsymbol{\hat{\theta}} \mid \boldsymbol{\mu}, \boldsymbol{\Sigma}_n + \boldsymbol{\mathcal{T}}) \times \frac{\Phi_K(\boldsymbol{0} \mid \boldsymbol{A}_{i,K} \boldsymbol{\mu}_*, \boldsymbol{A}_{i,K} \boldsymbol{\mathcal{T}}_* \boldsymbol{A}_{i,K}^\top)}{\Phi_K(\boldsymbol{0} \mid \boldsymbol{A}_{i,K}\boldsymbol{\mu}, \boldsymbol{A}_{i,K}\boldsymbol{\mathcal{T}}\boldsymbol{A}_{i,K}^\top)}
\end{align*}
where $\boldsymbol{A}_{i,K}$ is a $K \times K$ contrast matrix that maps
$\boldsymbol{\theta}$ to the space where the negative orthant corresponds to the
space of hypothesis $H_{+i}$. For example, for $i = 2$ and $K = 3$, the matrix
is
\begin{align*}
  \boldsymbol{A}_{2,3} =
  \begin{bmatrix}
    0 & -1 & 0 \\
    1 & -1 & 0 \\
    0 & -1 & 1 \\
  \end{bmatrix}
\end{align*}
with the first row encoding the constraint of $\theta_2$ being positive, and the
second and third rows encoding the constraints of $\theta_2$ being larger than
$\theta_1$ and $\theta_3$, respectively. As expected, for $K=1$, these marginal
likelihoods reduce to the univariate ones in the main text. Moreover, since they
are all available in closed form, Bayes factors, posterior probabilities, and
randomization probabilities are also available in closed form. Crucially, only
the exact evaluation of multivariate normal densities and cumulative
distribution functions is required, and both are efficiently implemented in
statistical software \citep[e.g., in the \texttt{mvtnorm} R
  package,][]{Genz2009}. This thus leads to an efficient Bayesian RAR method
that allows experimenters to compute randomization probabilities in complex
settings, for instance, multiple regression, where a full Bayesian analysis may
involve additional complexities, such as priors for nuisance parameters and
Markov chain Monte Carlo methods for the computation of posteriors.
\end{sloppypar}

\begin{figure}[p]
\begin{knitrout}
\definecolor{shadecolor}{rgb}{0.969, 0.969, 0.969}\color{fgcolor}

{\centering \includegraphics[width=\maxwidth]{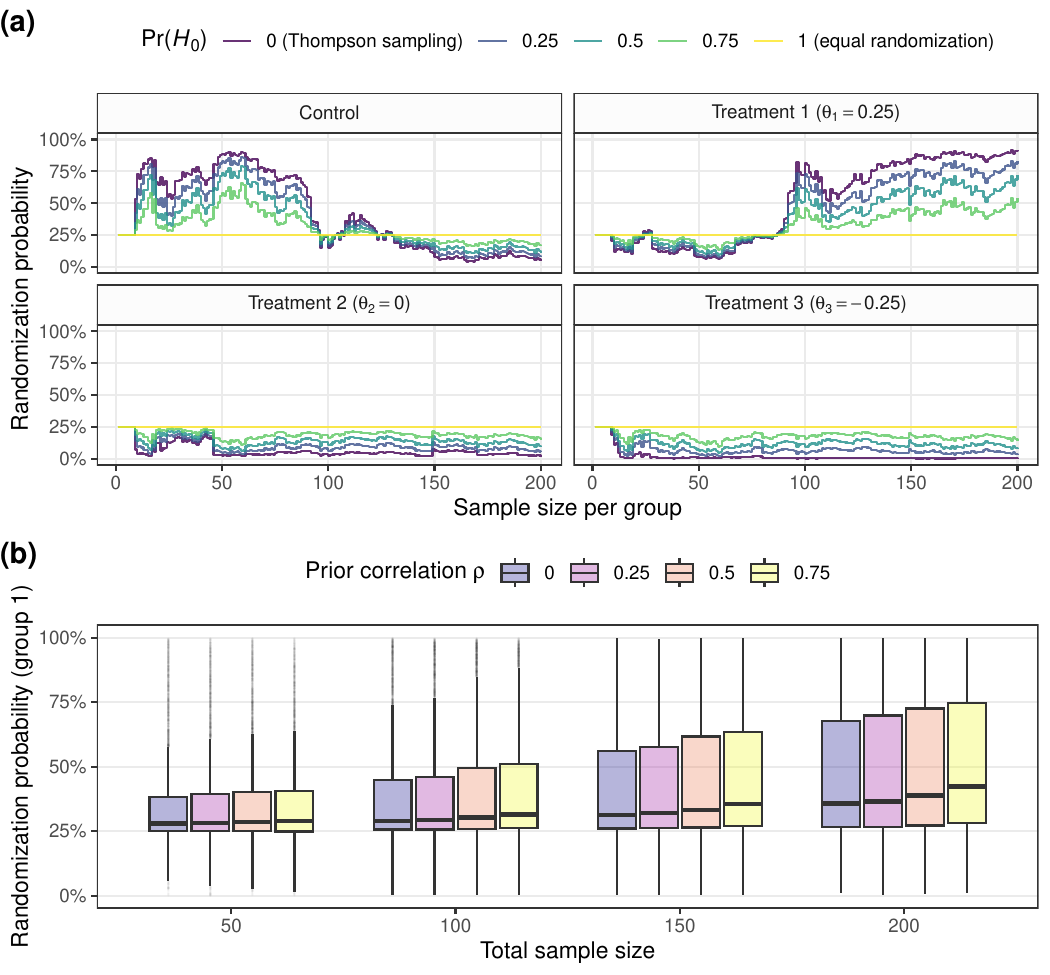} 

}

\end{knitrout}
\caption{Plot (a) shows a simulated evolution of Bayesian RAR probabilities for
  3 treatments and a control group. In each step, one patient is allocated to
  the control group or one of the treatment groups using Thompson sampling
  based on a normal prior centered at zero and with covariance matrix
  $\boldsymbol{\mathcal{T}}$ with $\boldsymbol{\mathcal{T}}_{ij} = 0.5$
  for $i\neq j$ and $\boldsymbol{\mathcal{T}}_{ij} = 1$ for $i = j$.
  Patient outcomes are then simulated from a normal distribution with a standard
  deviation of 1 and corresponding mean differences as indicated in the
  panel titles (treatment 1 is the most effective treatment). The mean
  difference effect estimate $\boldsymbol{\hat{\theta}}$ and its estimated
  covariance matrix $\boldsymbol{\Sigma}_n$ are computed with linear regression.
  Randomization probabilities are based on a spike-and-slab prior ($\Pr(H_0) >
  0$) with the same multivariate normal component as Thompson sampling, and are
  computed from the same data sequence to enable comparability. Plot (b) shows
  the distribution of randomization probabilities for different total sample
  sizes based on 10'000 simulations under
  Bayesian RAR with $\Pr(H_0) = 0.5$ and varying prior correlation
  $\boldsymbol{\mathcal{T}}_{ij} = \rho$ for $i\neq j$.}
\label{fig:example-multinormal}
\end{figure}

Figure~\ref{fig:example-multinormal}a shows sequences of randomization
probabilities computed from simulated normal data with $K = 3$ treatment
groups. We see that assigning a prior probability $\Pr(H_0) = 1$ leads to static
equal randomization at $\pi_i = 1/(K+1)= 25\%$,
while assigning $\Pr(H_0) = 0$ (Thompson sampling) leads to the most variable
randomization probabilities. Since data are simulated assuming that treatment 1
is the most effective, randomization probabilities based on $\Pr(H_0) < 1$
converge towards $\pi_1 = 100\%$ and toward $0\%$ for the remaining treatments.
While convergence is the fastest for $\Pr(H_0) = 0$, this prior probability also
accidentally produces rather high randomization probabilities for the control
group in the first half of the study, which is less pronounced for positive
prior probabilities $\Pr(H_0) > 0$.

\begin{sloppypar}
Figure~\ref{fig:example-multinormal}b shows the distribution of randomization
probabilities for the best treatment (treatment 1) simulated under Bayesian RAR
with $\Pr(H_0) = 0.5$ and for different correlation structures of the
multivariate normal prior distribution for the treatment effect vector. While
the simulation in Figure~\ref{fig:example-multinormal}a used $\rho = 0.5$ to
achieve equal prior probabilities of treatments at the study start, the
simulations in Figure~\ref{fig:example-multinormal}b also consider smaller and
larger values. We can see that with increasing correlation, the randomization
becomes more aggressive towards the effective treatment, though also the spread
of the randomization probabilities seems to increase.
\end{sloppypar}

\section{Simulation study}
\label{app:simulation}

We now describe the design and results of our simulation study following the
structured ADEMP approach \citep{Morris2019, Siepe2024}. Our simulation study
was not preregistered as it constitutes early-phase methodological research
where the properties of a new method are explored without the intention to give
wide recommendations for practitioners \citep{Heinze2023}. A website with
additional details and results is provided at
\url{https://samch93.github.io/brar/}.

\subsection{Aims}
The aim of the simulation study is to evaluate the design characteristics of the
newly proposed null hypothesis Bayesian RAR approach, and compare it to existing
methods.

\subsection{Data-generating mechanism}
The data-generating mechanism was inspired by the simulation studies from
\citet{Robertson2023}, \citet{Thall2007}, and \citet{Wathen2017}. In each
repetition, a data set with $n$ binary outcomes is simulated through RAR: A
patient $i$ is randomly allocated to the control group or one of the $K$
treatment groups based on randomization probabilities computed from the $1,
\dots, i -1$ preceding outcomes. Depending on the allocation, an outcome is
either simulated from a Bernoulli distribution with probability $\theta_C$ in
the control group, $\theta_1$ in the first treatment group, or $\theta_2$ for
the remaining treatment groups (in case $K > 1$).

Parameters were chosen similar to the simulation study from
\citet{Robertson2023}. We vary the sample size $n \in \{200, 654\}$ to represent
low- and high-powered studies, the number of treatment groups $K \in \{1, 2,
3\}$, and the probability in the first treatment group $\theta_1 \in \{0.25,
0.35, 0.45\}$. The probability in the control group and the remaining groups is
always fixed at $\theta_C = 0.25$ and $\theta_2 = \theta_3 = 0.3$, respectively.
All these parameters are varied fully factorially, leading to $2 \times 3 \times
3 = 18$ parameter conditions.

Since treatment allocation determines from which true probability an outcome is
simulated, data generation is directly influenced by the RAR methods described
below. These come with additional parameters that are, however, considered
method tuning parameters rather than true underlying parameters.

\subsection{Estimands and other targets}

The primary interest of this simulation study lies in assessing the patient
benefit characteristics of different RAR methods. Additionally, the estimand of
interest is the response rate difference $\text{RD}_1 = \theta_1 - \theta_C$ and
the target of interest is the null hypothesis of $\text{RD}_1 = 0$.

\subsection{Methods}

We consider the null hypothesis Bayesian RAR method described in the main text.
The prior probability of $H_0$ is a tuning parameter and controls the
variability of the randomization probabilities. Setting $\Pr(H_0) = 1$ produces
equal randomization, whereas $\Pr(H_0) = 0$ produces Thompson sampling. We
consider values of $\Pr(H_0) \in \{0, 0.25, 0.5, 0.75, 1\}$, as well as the
normal approximation and exact binomial versions of RAR (denoted ``Normal BRAR''
and ``Exact BRAR'' in the figures). Since $\Pr(H_0) = 1$ produces equal
randomization under both versions, it is simulated only with the exact version.
For approximate normal RAR, a normal prior with mean 0, variance 1, and in case
of $K > 1$ a correlation of 0.5, is considered. Independent uniform priors are
assigned for binomial RAR. Log odds ratios along with their covariance are
estimated with logistic regression and then used as inputs for the normal RAR
method, while the exact method uses success counts and sample sizes only. In
case a method fails to converge, equal randomization is applied as a back-up
strategy, as this mimics what an experimenter might do in practice when a RAR
method fails to converge \citep{Paweletal2025c}.

We also consider three modifications of these methods: In some conditions, a
``burn-in'' phase is carried out during which the first 50 patients are always
randomized with equal probability $1/(K + 1)$ to each group. For Thompson
sampling ($\Pr(H_0) = 0$), we additionally consider conditions with power
transformations of randomization probabilities, i.e., if $\pi_k$ is the
randomization probability of group $k$, we take $\pi_k^* = \pi_k^c / \sum_{j\in
  \{C,1,\dots,K\}} \pi_j^c$. We consider $c=1/2$ and $c=i/(2n)$ with $i$ the
current and $n$ the maximal sample size, which are two popular choices of the
tuning parameter $c$ \citep{Wathen2017}. Additionally, in some conditions
``capping'' is applied to Thompson sampling. That is, randomization
probabilities outside the $[10\%, 90\%]$ interval are set to either 10\% or
90\%. After capping has been performed, randomization probabilities are
re-normalized to sum to one \citep{Wathen2017, Kim2021}. This re-normalization
is only performed for randomization probabilities greater than 10\%, as these
would otherwise be reduced again to probabilities less than 10\%. In case a
re-normalized probability becomes less than 10\%, it is also capped at 10\% and
a second re-normalization performed. For equal randomization ($\Pr(H_0) = 1$) no
burn-in is simulated as this manipulation has no effect.

Finally, we also include two methods from the multi-armed bandit literature in
the simulation study to compare null hypothesis Bayesian RAR to more aggressive
allocation methods -- the Gittins index \citep{Gittins2011, Villar2015} and the
Bayesian upper confidence bound (Bayes UCB) methods \citep{Kaufmann2012}. For
the Gittins index, we use the implementation from the \texttt{RARtrials} R
package \citep{Xu2025} with a discount factor of $d = 0.995$. For Bayes UCB we
use uniform priors for each probability along with the tuning parameter $c = 0$,
as recommended in \citet{Kaufmann2012} based on their simulation results.

\subsection{Performance measures}

Patient benefit was quantified with:
\begin{itemize}
\item The mean rate of successes per study
  \begin{align*}
    \overline{\text{RS}} = \frac{1}{n_{\text{sim}}} \sum_{i=1}^{n_{\text{sim}}} \, \sum_{j=1}^n
  \frac{y_{ij}}{n}
  \end{align*}
  where $y_{ij}$ denotes the 0/1 success indicator of patient $j$ in simulation
  $i$, $n$ is the sample size, and $n_{\text{sim}}$ is the number of simulation
  repetitions.

  \item The mean rate of extreme randomization probabilities (less than 10\% or
    greater than 90\%)
  \begin{align*}
    \overline{\text{REP}} = \frac{1}{n_{\text{sim}}} \sum_{i=1}^{n_{\text{sim}}} \sum_{j=1}^n \frac{\mathbb{1}(\text{any randomization probability at time } j ~ < 10\% ~ \text{or} ~ > 90\%)}{n}
  \end{align*}
  with indicator function $\mathbb{1}(\cdot)$.

\item The proportion of simulations in which the average sample size of the
  other groups exceeded that of treatment group 1 by more than 10\% of the total
  sample size
  \begin{align*}
  \hat{S}_{0.1} = \frac{1}{n_{\text{sim}}} \sum_{i=1}^{n_{\text{sim}}} \,
  \mathbb{1}\left(\frac{n - n_{1i}}{K} - n_{1i} > 0.1n\right)
  \end{align*}
  where $n_{1i}$ is the number of allocations to treatment group 1 in simulation
  $i$. For $K= 1$, this reduces to the $\hat{S}_{0.1}$ sample size imbalance measure
  from \citet{Robertson2023}, which in turn was inspired by the performance
  evaluation in the simulation study from \citet{Thall2015}. Note that the
  $\hat{S}_{0.1}$ measure is difficult to translate into an actual patient
  benefit measure because the patient benefit loss related to imbalance depends
  on the size of the treatment effect \citep{Robertson2023}. For instance,
  imbalance is less consequential if the treatment effect is small than if it is
  large. Nevertheless, we include the $\hat{S}_{0.1}$ measure for comparability
  with previous simulation studies.

\end{itemize}
Parameter estimation and hypothesis testing performance were quantified with:
\begin{itemize}

\item The empirical bias of the estimate of the response rate difference $\text{RD}_1$
  \begin{align*}
  \mathrm{Bias}(\text{RD}_1) = \frac{1}{n_{\text{sim}}} \sum_{i=1}^{n_{\text{sim}}} \,
  \hat{\theta}_{1i} - \hat{\theta}_{Ci} - \text{RD}_1
  \end{align*}
  where $\hat{\theta}_{1i}$ and $\hat{\theta}_{Ci}$ are the maximum likelihood
  estimates of the probabilities $\theta_1$ and $\theta_C$ in simulation
  repetition $i$.

\item Empirical coverage of the $95\%$ Wald confidence intervals of
  $\text{RD}_1$
  \begin{align*}
  \mathrm{Coverage}(\text{RD}_1) = \frac{1}{n_{\text{sim}}} \sum_{i=1}^{n_{\text{sim}}} \,
  \mathbb{1}\left\{95\% ~ \text{CI includes} ~ \text{RD}_1 ~ \text{in simulation} ~ i\right\}.
  \end{align*}

\item Empirical rejection rate (type I error rate or power depending on the
  condition) related to the Wald test of the response rate difference $\text{RD}_1$
  \begin{align*}
  \mathrm{RR}(\text{RD}_1) = \frac{1}{n_{\text{sim}}} \sum_{i=1}^{n_{\text{sim}}} \,
  \mathbb{1}\left\{\text{Test rejects} ~ H_0 \colon \text{RD}_1 = 0 ~ \text{in simulation} ~ i\right\}.
  \end{align*}
\end{itemize}

Each condition was simulated 10'000 times. This
ensures an MCSE (Monte Carlo Standard Error) for type I error rate, power, and
coverage of at most 0.5\%. MCSEs were calculated using
the formulae from \citet{Siepe2024} and are provided for all measures in the
following figures and the supplemental website.

\subsection{Computational aspects}

The simulation study was run on a server running Debian GNU/Linux forky/sid and
R version 4.5.2 (2025-10-31). The SimDesign R package was used to
organize and run the simulation study \citep{Chalmers2020}. Our newly developed
\texttt{brar} R package was used to perform Bayesian RAR. Code and data to
reproduce this simulation study are available at
\url{https://github.com/SamCH93/brar} and persistently archived at
\url{https://doi.org/10.5281/zenodo.17248628}.

\subsection{Results}

\paragraph{Convergence}
\begin{sloppypar}
Non-convergence happened only rarely for the normal approximation method due to
logistic regression not converging at the start of the study when no events were
observed for some groups. In this case, equal randomization was applied. The
highest rate of such non-convergence was in a condition with $K = 3$ treatment
groups where 4.6\% of the $n$
logistic regressions did not converge. No other forms of missingness were
observed. Figures and tables with per-condition-method non-convergence rates are
available at \url{https://samch93.github.io/brar/}.
\end{sloppypar}

\afterpage{
\begin{landscape}

\begin{figure}[!htb]
\begin{knitrout}
\definecolor{shadecolor}{rgb}{0.969, 0.969, 0.969}\color{fgcolor}

{\centering \includegraphics[width=\maxwidth]{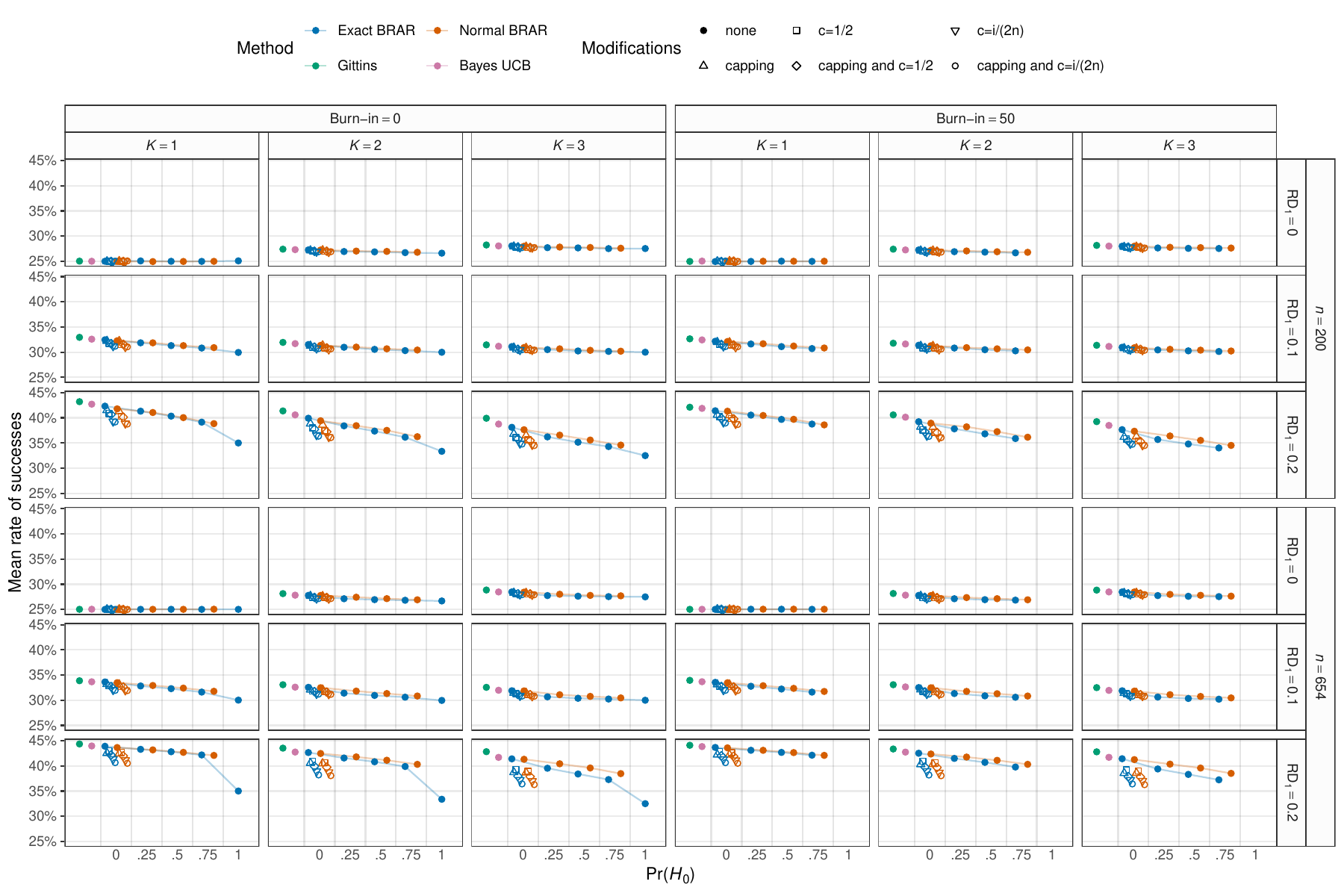} 

}

\end{knitrout}
\caption{Mean rate of successes (i.e., the number of successes in a study
  divided by its sample size averaged over all 10'000 simulation repetitions). The maximum MCSE is
  0.051\%. Gittins index and Bayes
  UCB do not depend on $\Pr(H_0)$ and are shown to the left of $\Pr(H_0) = 0$
  for comparison.}
\label{fig:successrate}
\end{figure}

\begin{figure}[!htb]
\begin{knitrout}
\definecolor{shadecolor}{rgb}{0.969, 0.969, 0.969}\color{fgcolor}

{\centering \includegraphics[width=\maxwidth]{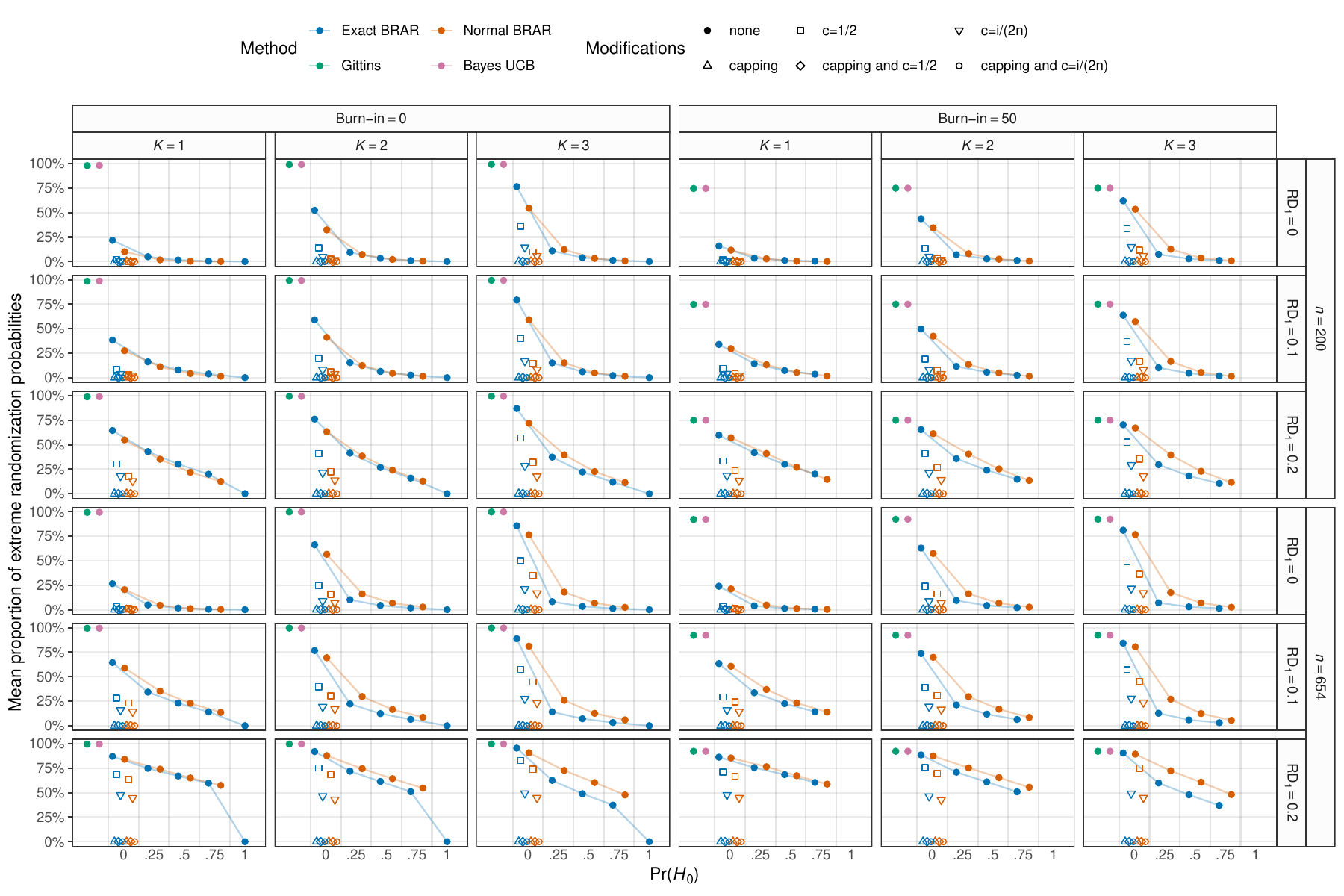} 

}

\end{knitrout}
\caption{Mean proportion of extreme randomization probabilities (either less
  than 10\% or greater than 90\%) averaged over all 10'000 simulation repetitions. The maximum MCSE is
  0.33\%. Gittins index and
  Bayes UCB do not depend on $\Pr(H_0)$ and are shown to the left of $\Pr(H_0) =
  0$ for comparison.}
\label{fig:extreme}
\end{figure}

\begin{figure}[!htb]
\begin{knitrout}
\definecolor{shadecolor}{rgb}{0.969, 0.969, 0.969}\color{fgcolor}

{\centering \includegraphics[width=\maxwidth]{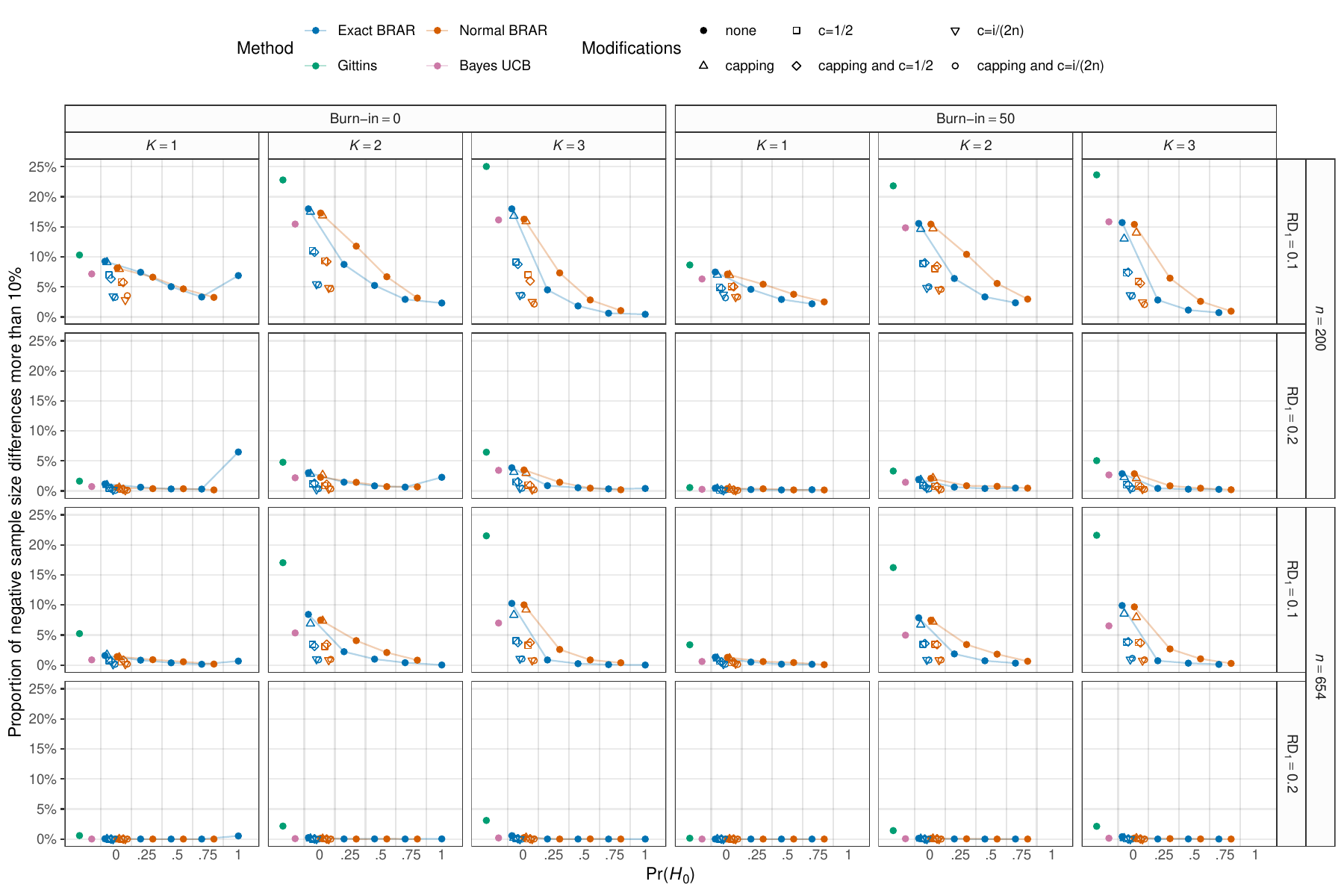} 

}

\end{knitrout}
\caption{Proportion of 10'000 simulations in which
  the average sample size of the other groups exceeded the sample size of
  treatment group 1 by more than 10\% of the total sample size. The maximum MCSE
  is 0.5\%. Gittins index and
  Bayes UCB do not depend on $\Pr(H_0)$ and are shown to the left of $\Pr(H_0) =
  0$ for comparison.}
\label{fig:imbalance}
\end{figure}

\begin{figure}[!htb]
\begin{knitrout}
\definecolor{shadecolor}{rgb}{0.969, 0.969, 0.969}\color{fgcolor}

{\centering \includegraphics[width=\maxwidth]{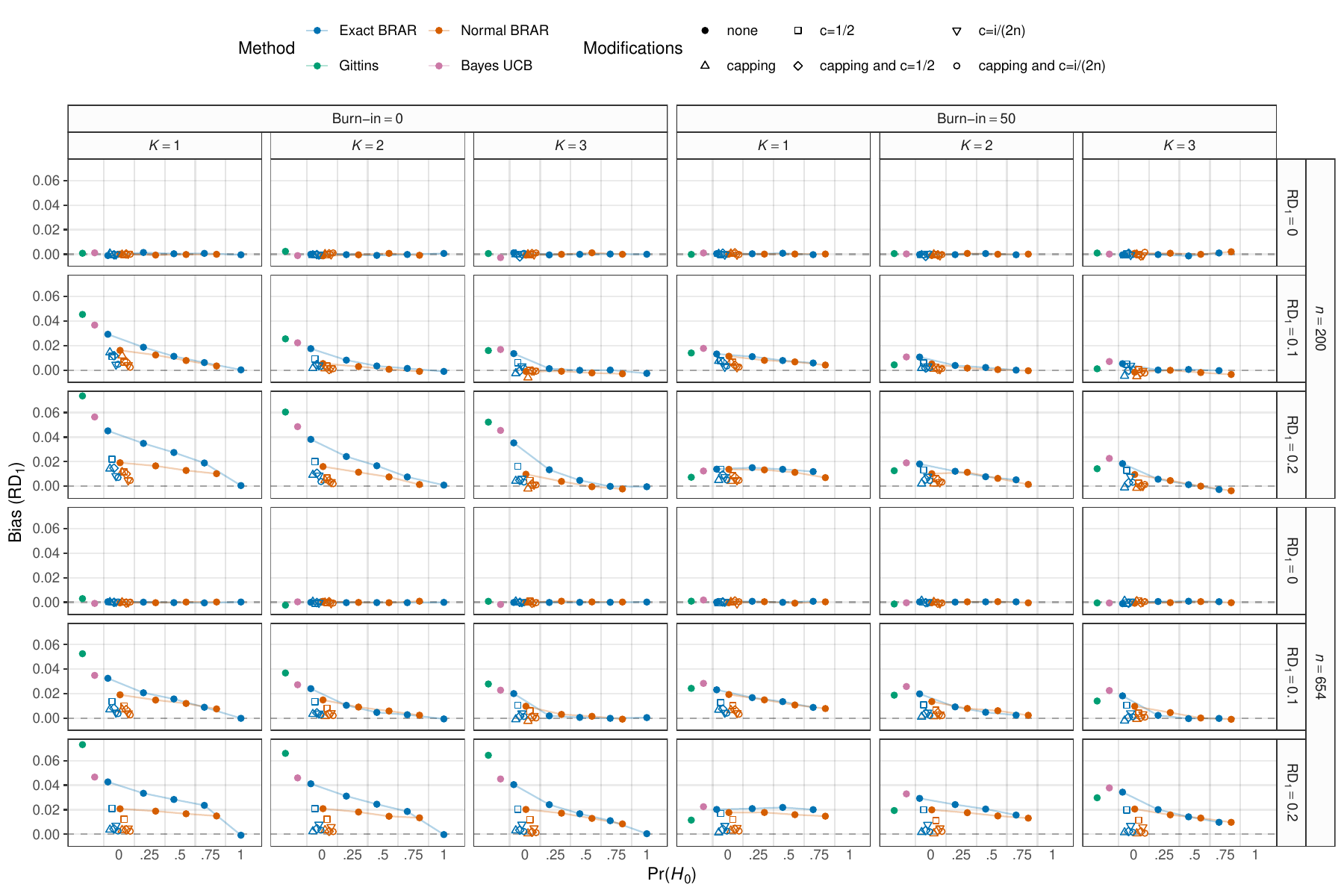} 

}

\end{knitrout}
\caption{Empirical bias of the estimate of the response rate difference
  $\text{RD}_1$ between the first treatment group and the control group based on
  10'000 simulation repetitions. The maximum MCSE
  is 0.0016. Gittins index and Bayes
  UCB do not depend on $\Pr(H_0)$ and are shown to the left of $\Pr(H_0) = 0$
  for comparison.}
\label{fig:bias}
\end{figure}

\begin{figure}[!htb]
\begin{knitrout}
\definecolor{shadecolor}{rgb}{0.969, 0.969, 0.969}\color{fgcolor}

{\centering \includegraphics[width=\maxwidth]{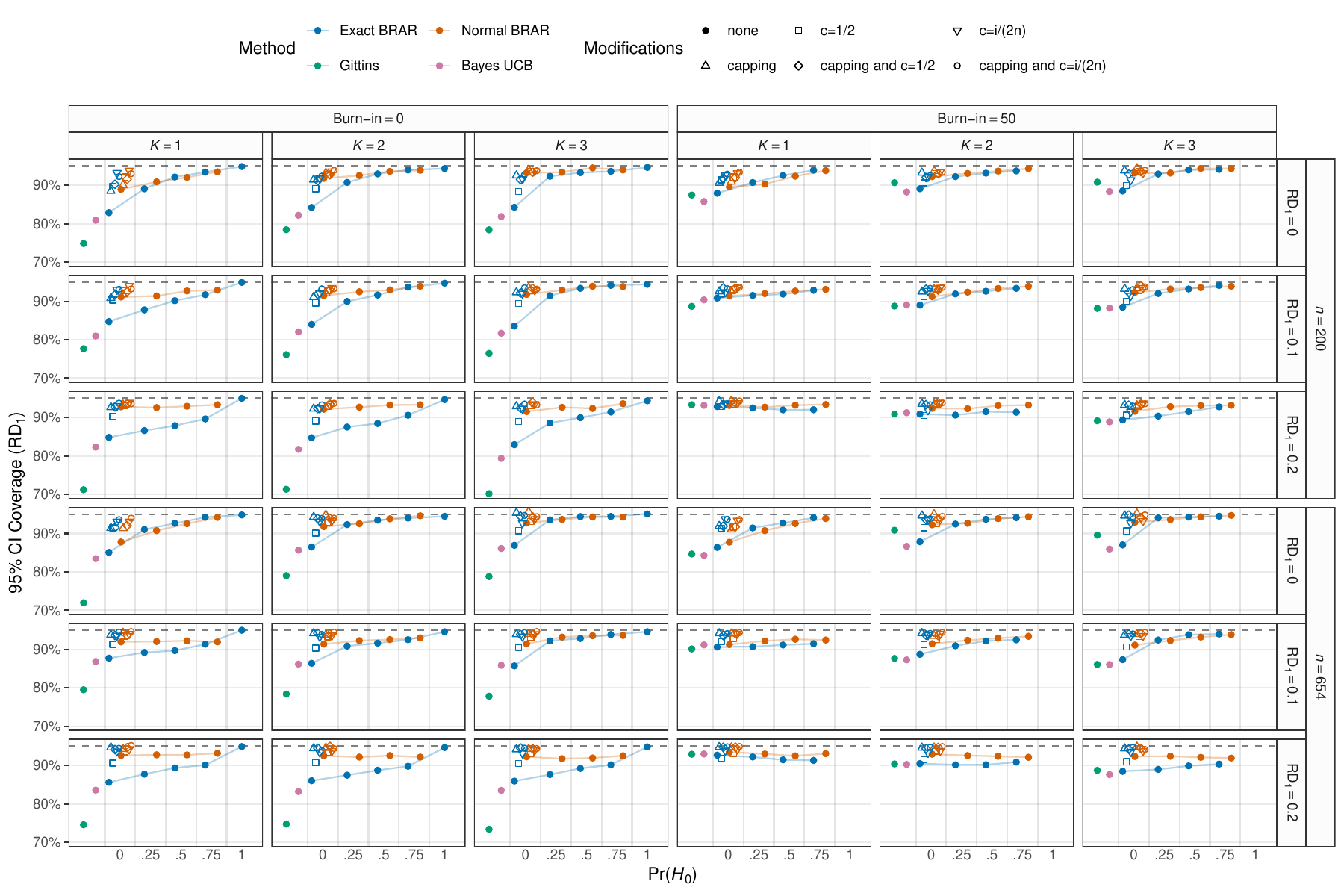} 

}

\end{knitrout}
\caption{Empirical coverage of the 95\% Wald confidence interval for the
  response rate difference $\text{RD}_1$ based on 10'000 simulation repetitions. The maximum MCSE is
  0.46\%. Gittins index and
  Bayes UCB do not depend on $\Pr(H_0)$ and are shown to the left of $\Pr(H_0) =
  0$ for comparison.}
\label{fig:coverage}
\end{figure}

\begin{figure}[!htb]
\begin{knitrout}
\definecolor{shadecolor}{rgb}{0.969, 0.969, 0.969}\color{fgcolor}

{\centering \includegraphics[width=\maxwidth]{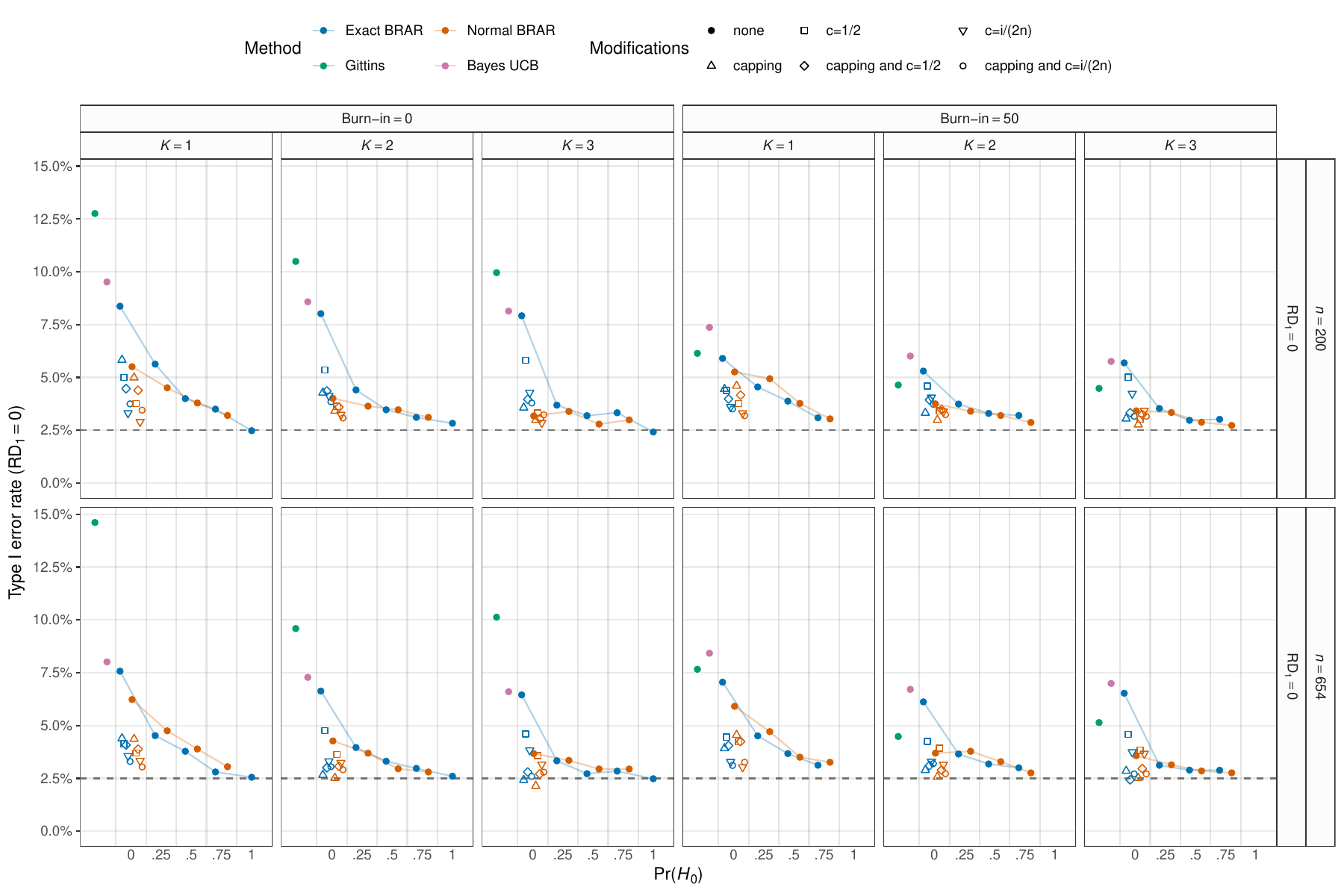} 

}

\end{knitrout}
\caption{Empirical type I error rate of the Wald test of $\text{RD}_1 = 0$ based
  on 10'000 simulation repetitions. The maximum
  MCSE is 0.35\%. Gittins index and Bayes UCB do not depend on $\Pr(H_0)$ and are shown
  to the left of $\Pr(H_0) = 0$ for comparison.}
\label{fig:t1e}
\end{figure}

\begin{figure}[!htb]
\begin{knitrout}
\definecolor{shadecolor}{rgb}{0.969, 0.969, 0.969}\color{fgcolor}

{\centering \includegraphics[width=\maxwidth]{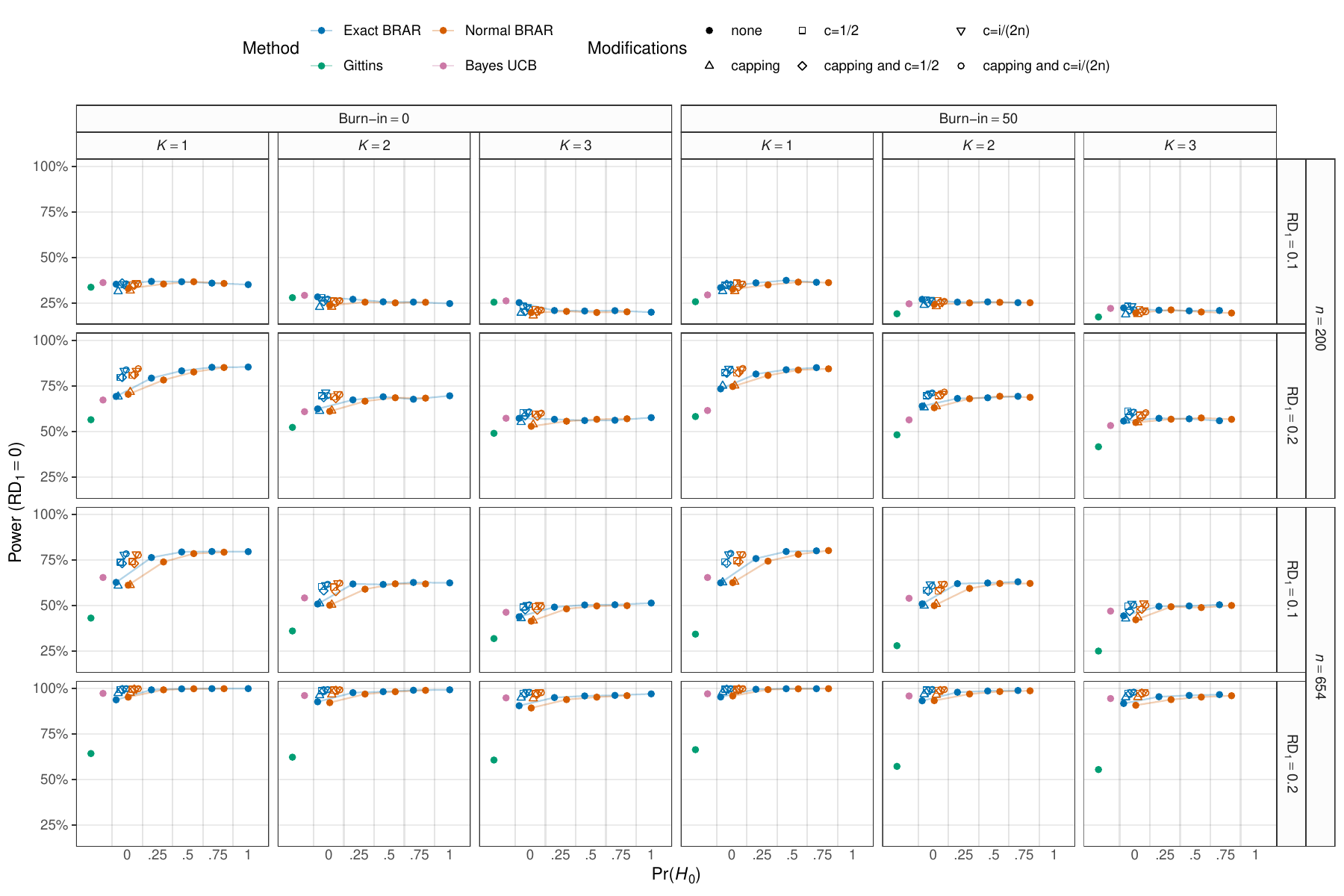} 

}

\end{knitrout}
\caption{Empirical power of the Wald test of $\text{RD}_1 = 0$ based on
  10'000 simulation repetitions. The maximum MCSE
  is 0.5\%.
  Gittins index and Bayes UCB do not depend on $\Pr(H_0)$ and are shown to the
  left of $\Pr(H_0) = 0$ for comparison.}
\label{fig:power}
\end{figure}

\end{landscape}
}

\paragraph{Rate of successes and extreme randomization probabilities}
The mean rate of successes is shown in Figure~\ref{fig:successrate}. It was
generally the highest for the Gittins index, followed by Bayes UCB, Thompson
sampling, and was lowest for equal randomization. The normal approximation and
exact method produced mostly similar rates, with the normal method sometimes
showing slightly higher rates (e.g., for $K = 3$ and $\text{RD}_1 = 0.2$). The
Thompson sampling modifications generally reduced the mean success rate, with
the greatest reduction achieved by combining capping and power transformation
with $c = i/(2n)$. In conditions with small sample size these rates were similar
to when the prior probability was $\Pr(H_0) = 0.75$, while they were lower when
the sample size was larger. This makes sense as for larger sample sizes,
uncapped randomization probabilities are more likely to converge to extreme
ones. This is also visible in Figure~\ref{fig:extreme} where more extreme
randomization probabilities are observed with increasing sample size and
response rate difference for all methods but the ones with capping. Finally,
Bayes UCB and the Gittins index methods always produced extreme randomization
probabilities, as these methods deterministically allocate the next patient to
the group with the highest upper confidence bound and Gittins index,
respectively.

\paragraph{Negative sample size imbalance}
Figure~\ref{fig:imbalance} shows negative sample size imbalance as quantified by
the $\hat{S}_{0.1}$ metric. The Gittins index generally showed the highest
negative imbalance, followed by Thompson sampling and Bayes UCB. Among the
methods with $\Pr(H_0) < 1$, negative imbalance was the greatest for Thompson
sampling and reduced when modifications were applied. Similarly, increasing the
prior probability of $H_0$ decreased negative imbalance, in some cases even
below Thompson sampling with modifications. As in the simulation study from
\citet{Robertson2023}, negative imbalance of Thompson sampling and Bayesian RAR
with $\Pr(H_0) < 1$ was mostly higher than under equal randomization in conditions with
small effect size $\text{RD}_1 = 0.1$, but equal or lower than under equal
randomization in conditions with large effect size $\text{RD}_1 = 0.2$.

\paragraph{Bias and coverage}
Figures~\ref{fig:bias} and~\ref{fig:coverage} show empirical bias and coverage
related to estimates of the response rate difference between the first treatment
group and the control group $\text{RD}_1$. In conditions where there was no
difference ($\text{RD}_1 = 0$), all methods produced unbiased point estimates,
though all methods but equal randomization also showed undercoverage. Bias
occurred in conditions with non-zero rate differences, the bias being the
greatest for the Gittins index, Bayes UCB, and Thompson sampling, and decreasing
to some extent when modifications were introduced or the prior probability of
$H_0$ increased. Similarly, modifications or increasing prior probabilities
improved coverage, although coverage still remained suboptimal in most
conditions. For large effect size conditions, coverage was often
better for the normal than the exact version of RAR. Similarly, bias was in some
cases somewhat larger for the exact compared to the normal version. This could
potentially be due to the normal prior distribution on the log odds ratio being
better tuned to effect sizes in the data-generating mechanism compared to the
uniform priors on the probabilities under the binomial model.

\paragraph{Type I error rate and power}
Figures~\ref{fig:t1e} and~\ref{fig:power} show empirical type I error rate and
power associated with the Wald test of $\text{RD}_1 = 0$. We see that the
Gittins index, Bayes UCB, and Thompson sampling methods showed an inflated type
I error rate above the nominal 2.5\% which was reduced to some extent by
increasing either the prior probability of $H_0$ or applying modifications. In
the same way, the Gittins index, Bayes UCB, and Thompson sampling often
exhibited reduced power compared to equal randomization, which was again
alleviated by modifications or increasing $\Pr(H_0)$. For small sample sizes and
large rate differences, power was slightly increased for the exact compared to
the normal version of RAR, while for Thompson sampling the type I error rate was
slightly higher for the exact compared to the normal version.

\section{The R package brar}
\label{app:package}
Our R package can be installed by running \texttt{install.packages("brar")} in
an R console. The main functions of the package are \texttt{brar\_normal} and
\texttt{brar\_binomial}, which implement the normal method from Section~3 and
the binomial method from Section~4 in the main text. The following code chunk
illustrates how the latter function can be used.

\begin{spacing}{1}
\begin{knitrout}
\definecolor{shadecolor}{rgb}{0.969, 0.969, 0.969}\color{fgcolor}\begin{kframe}
\begin{alltt}
\hlkwd{library}\hldef{(brar)} \hlcom{# load package}

\hlcom{## successes and observations in control and 3 treatment groups}
\hldef{y} \hlkwb{<-} \hlkwd{c}\hldef{(}\hlnum{10}\hldef{,} \hlnum{9}\hldef{,} \hlnum{14}\hldef{,} \hlnum{13}\hldef{)}
\hldef{n} \hlkwb{<-} \hlkwd{c}\hldef{(}\hlnum{20}\hldef{,} \hlnum{20}\hldef{,} \hlnum{22}\hldef{,} \hlnum{21}\hldef{)}

\hlcom{## conduct exact point null Bayesian RAR}
\hldef{res} \hlkwb{<-} \hlkwd{brar_binomial}\hldef{(}\hlkwc{y} \hldef{= y,} \hlkwc{n} \hldef{= n,}
                     \hlcom{## uniform prior for common probability under H0}
                     \hlkwc{a0} \hldef{=} \hlnum{1}\hldef{,} \hlkwc{b0} \hldef{=} \hlnum{1}\hldef{,}
                     \hlcom{## uniform priors for all probabilities}
                     \hlkwc{a} \hldef{=} \hlkwd{c}\hldef{(}\hlnum{1}\hldef{,} \hlnum{1}\hldef{,} \hlnum{1}\hldef{,} \hlnum{1}\hldef{),} \hlkwc{b} \hldef{=} \hlkwd{c}\hldef{(}\hlnum{1}\hldef{,} \hlnum{1}\hldef{,} \hlnum{1}\hldef{,} \hlnum{1}\hldef{),}
                     \hlcom{## prior probability of the null hypothesis}
                     \hlkwc{pH0} \hldef{=} \hlnum{0.5}\hldef{)}
\hldef{res} \hlcom{# print results}
\end{alltt}
\begin{verbatim}
## DATA
##             Events  N Proportion
## Control         10 20      0.500
## Treatment 1      9 20      0.450
## Treatment 2     14 22      0.636
## Treatment 3     13 21      0.619
## 
## PRIOR PROBABILITIES
##    H-    H0   H+1   H+2   H+3 
## 0.125 0.500 0.125 0.125 0.125 
## 
## BAYES FACTORS (BF_ij)
##         H-     H0   H+1    H+2   H+3
## H-   1.000 0.0341  2.16 0.1837 0.223
## H0  29.335 1.0000 63.45 5.3891 6.533
## H+1  0.462 0.0158  1.00 0.0849 0.103
## H+2  5.443 0.1856 11.77 1.0000 1.212
## H+3  4.490 0.1531  9.71 0.8249 1.000
## 
## POSTERIOR PROBABILITIES
##      H-      H0     H+1     H+2     H+3 
## 0.00777 0.91148 0.00359 0.04228 0.03488 
## 
## RANDOMIZATION PROBABILITIES
##     Control Treatment 1 Treatment 2 Treatment 3 
##       0.236       0.231       0.270       0.263
\end{verbatim}
\begin{alltt}
\hldef{res}\hlopt{$}\hldef{prand} \hlcom{# extract randomization probabilities for RAR}
\end{alltt}
\begin{verbatim}
##     Control Treatment 1 Treatment 2 Treatment 3 
##   0.2356374   0.2314611   0.2701530   0.2627485
\end{verbatim}
\end{kframe}
\end{knitrout}
\end{spacing}

\section*{Software and data}
Code and data to reproduce our analyses are openly available at
\url{https://github.com/SamCH93/brar}. A snapshot of the repository at the time
of writing is available at \url{https://doi.org/10.5281/zenodo.17248628}. We
used the statistical programming language
R version 4.6.1 (2026-06-24) for analyses \citep{R2026} along with
the \texttt{ggplot2} \citep{Wickham2016}, \texttt{dplyr} \citep{Wickham2023},
\texttt{SimDesign} \citep{Chalmers2020}, \texttt{mvtnorm} \citep{Genz2009},
\texttt{ggh4x} \citep{Brand2024}, \texttt{ggpubr} \citep{Kassambara2023},
\texttt{RARtrials} \citep{Xu2025}, and \texttt{knitr} \citep{Xie2015} packages.
\end{appendices}

\end{document}